\documentclass[preprints,article,accept,pdftex,moreauthors]{Definitions/mdpi}
\firstpage{1}
\pubvolume{1}
\issuenum{1}
\articlenumber{0}
\pubyear{}
\copyrightyear{}
\externaleditor{} 
\datereceived{}
\daterevised{} 
\dateaccepted{}
\datepublished{}
\hreflink{\textls[-25]{https://doi.org/}} 

\newcommand{\xadded}[1]{#1}
\newcommand{\xdeleted}[1]{}
\newcommand{\xreplaced}[2]{#1}

\Title{Electronic Structure and Minimal Models for Flat and Corrugated $\mathrm{CuO}$ Monolayers: An \text{Ab Initio} Study} 

\TitleCitation{Electronic Structure and Minimal Models for Flat and Corrugated $\mathrm{CuO}$ Monolayers: An \text{Ab Initio} Study} 

\Author{{Anatoly A. Slobodchikov} 
$^{1,}$*\orcidA{}, {Igor} 
A. Nekrasov $^{1}$\orcidB{}, {Lyudmila} V. Begunovich $^{2,3}$\orcidF{}, {Ilya} 
A. Makarov $^{3}$\orcidC{}, \linebreak {Maxim} M. Korshunov $^{3}$\orcidD{} and {Sergey} G. Ovchinnikov $^{3}$\orcidE{}}

\AuthorNames{Anatoly A. Slobodchikov, Igor A. Nekrasov, Lyudmila V. Begunovich, Ilya A. Makarov,  Maxim M. Korshunov and Sergey G. Ovchinnikov}

\AuthorCitation{Slobodchikov, A.A.; Nekrasov, I.A.; Begunovich, L.V.; Makarov, I.A.; Korshunov, M.M.; Ovchinnikov, S.G.}

\address{%
$^{1}$ \quad Institute of Electrophysics, Russian Academy of Sciences, Ural Branch, 620016 Yekaterinburg, Russia; nekrasov@iep.uran.ru (I.A.N.)\\
$^{2}$ \quad {Federal Research Center KSC SB RAS,} 
 Akademgorodok, 660036 Krasnoyarsk, Russia; beg@iph.krasn.ru\\
$^{3}$ \quad Kirensky Institute of Physics, Federal Research Center KSC SB RAS, Akademgorodok, \linebreak 660036 Krasnoyarsk, Russia; maki@iph.krasn.ru (I.A.M.); mkor@iph.krasn.ru (M.M.K.); \linebreak sgo@iph.krasn.ru (S.G.O.)}

\corres{Correspondence: slobodchikov@iep.uran.ru}

\abstract{$\mathrm{CuO}$ atomic thin monolayer ($\mathrm{mlCuO}$) was synthesized recently. Interest in the $\mathrm{mlCuO}$ is based on its close relation to $\mathrm{CuO_2}$ layers in typical high temperature cuprate superconductors. Here, we present the calculation of the band structure, the density of states and the Fermi surface of the flat $\mathrm{mlCuO}$ as well as the corrugated $\mathrm{mlCuO}$ within the density functional theory (DFT) in the generalized gradient approximation (GGA). In the flat $\mathrm{mlCuO}$, the $\mathrm{Cu}$-$3d_{x^2-y^2}$ band crosses the Fermi level, while the $\mathrm{Cu}$-$3d_{xz,yz}$ hybridized band is located just below it. The corrugation leads to a significant shift of the $\mathrm{Cu}$-$3d_{xz,yz}$ hybridized band down in energy and \xadded{a degeneracy lifting for the $\mathrm{Cu}$-$3d_{x^2-y^2}$ bands.} Corrugated $\mathrm{mlCuO}$ is more energetically favorable than the flat one. In addition, we compared the electronic structure of the considered $\mathrm{CuO}$ monolayers with bulk $\mathrm{CuO}$ systems. We also investigated the influence of a crystal lattice strain (which might occur on some interfaces) on the electronic structure of both $\mathrm{mlCuO}$ and determined the critical strains of topological Lifshitz transitions. Finally, we proposed a number of different minimal models for the flat and the corrugated $\mathrm{mlCuO}$ using projections onto different Wannier functions basis sets and obtained the corresponding Hamiltonian matrix elements in a real space.}


\begin{document}

\section{Introduction}

Copper oxides stand apart from other transition metal compounds. First of all, they attract much attention because of their high temperature superconductivity (HTSC) and already existing applications~\cite{anisimov_band_1991, ruiz_electronic_1997,ghijsen_electronic_1988,heinemann_band_2013,pickett_electronic_1989} as catalysts~\cite{reitz_propylene_1998}, photocells~\cite{nakaoka_photoelectrochemical_2004} and thin-film transistors~\cite{kim_p-channel_2013}. $\mathrm{CuO}$ is an exceptional member of the generally rocksalt family ($\mathrm{MnO}$ to $\mathrm{CuO}$), as it deviates both structurally and electronically from others. Unlike other members of the $3d$ transition oxides, which crystallize in the cubic rocksalt structure (with possible rhombohedral distortions), Tenorite (CuO) crystallizes in the lower symmetry monoclinic (\textit{C2/c}) crystal structure~\cite{asbrink_refinement_1970}, albeit the cubic crystal structure is also possible~\cite{schmahl_uber_1968}. Thus far, the bulk compound $\mathrm{CuO}$ has been thoroughly studied using \text{ab initio} calculations: DFT+U~\cite{ekuma_electronic_2014,wu_mathrmlsdamathrmu_2006,nolan_p-type_2006,cao_dft_2018,grant_electronic_2008}, DFT with hybrid functional~\cite{heinemann_band_2013} and Charge Transition Level Approach~\cite{cipriano_band_2020}.

The relatively recent interest in the $\mathrm{CuO}$ monolayer arose in part because one would expect a superconducting phase to occur here by analogy with the typical representative of HTSC cuprates $\mathrm{La_2CuO_4}$. There, the superconductivity occurs in two-dimensional layers formed by $\mathrm{CuO_2}$ plaquettes. The $\mathrm{CuO}$ monolayer consists of the same plaquettes. However, the plaquettes in the cuprates are connected by vertices, whereas in the monolayer they are connected by faces. This fact leads to a difference in the chemical composition---the number of copper and oxygen atoms is identical in the monolayer ($\mathrm{CuO}$), while there are two oxygen atoms per copper atom in the cuprates ($\mathrm{CuO_2}$).

In general, the electronic properties of the copper oxides are thoroughly studied. As everyone knows, $\mathrm{La_2CuO_4}$ has the $\mathrm{Cu}$-$d_{x^2-y^2}$ orbital at the Fermi level~\cite{pickett_electronic_1989}. In Ref.~\cite{yazdani_first-principles_2021} the authors studied isolated $\mathrm{CuO_2}$ monolayer using DFT and showed that all $\mathrm{Cu}$ orbitals, except for $d_{z^2}$, have states near the Fermi level; the $\mathrm{Cu}$-$d_{xz}$ and $d_{yz}$ orbitals and the $\mathrm{O}$-$p_z$ orbital have most of the states near the Fermi level, leading to $\pi$ bonds in the entire monolayer. The bulk $\mathrm{CuO}$ with the monoclinic structure, known as a $p$-type semiconductor, has significantly three dimensional electronic structure with mainly $\mathrm{Cu}$-$3d$ ($\mathrm{Cu}$-$3d_{x^2-y^2}$) states near the Fermi level~\cite{ekuma_electronic_2014}. DFT study in Ref.~\cite{cao_dft_2018} claims that the bulk $\mathrm{CuO}$ with the cubic structure is an indirect gap semiconductor; its valence band consists mainly of $\mathrm{O}$-$2p$ and $\mathrm{Cu}$-$3d$ orbitals.
\xadded{We can conclude that the electronic structure of the $\mathrm{CuO_2}$ monolayer and $\mathrm{La_2CuO_4}$ is quite similar to each other, while the one of the flat $\mathrm{mlCuO}$ partially resembles them, but has some qualitatively differences, such as an extremum presence in the $\Gamma - M$ direction (as the reader can observe later).}

\begin{figure}[H]
\begin{minipage}[h]{0.23\linewidth}
\center{\includegraphics[width=0.75\linewidth]{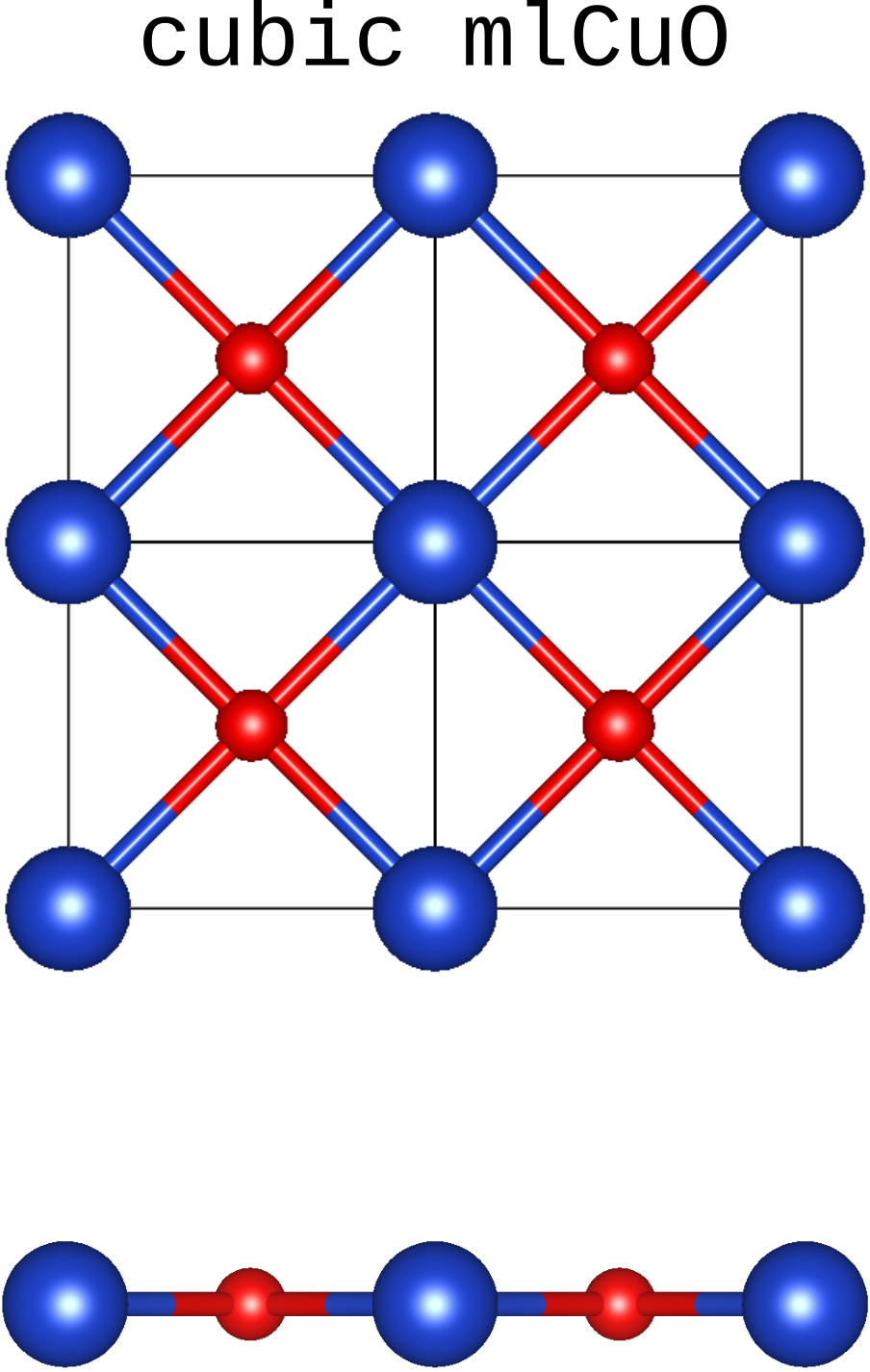}}
\end{minipage}
\hfill
\begin{minipage}[h]{0.23\linewidth}
\center{\includegraphics[width=0.93\linewidth]{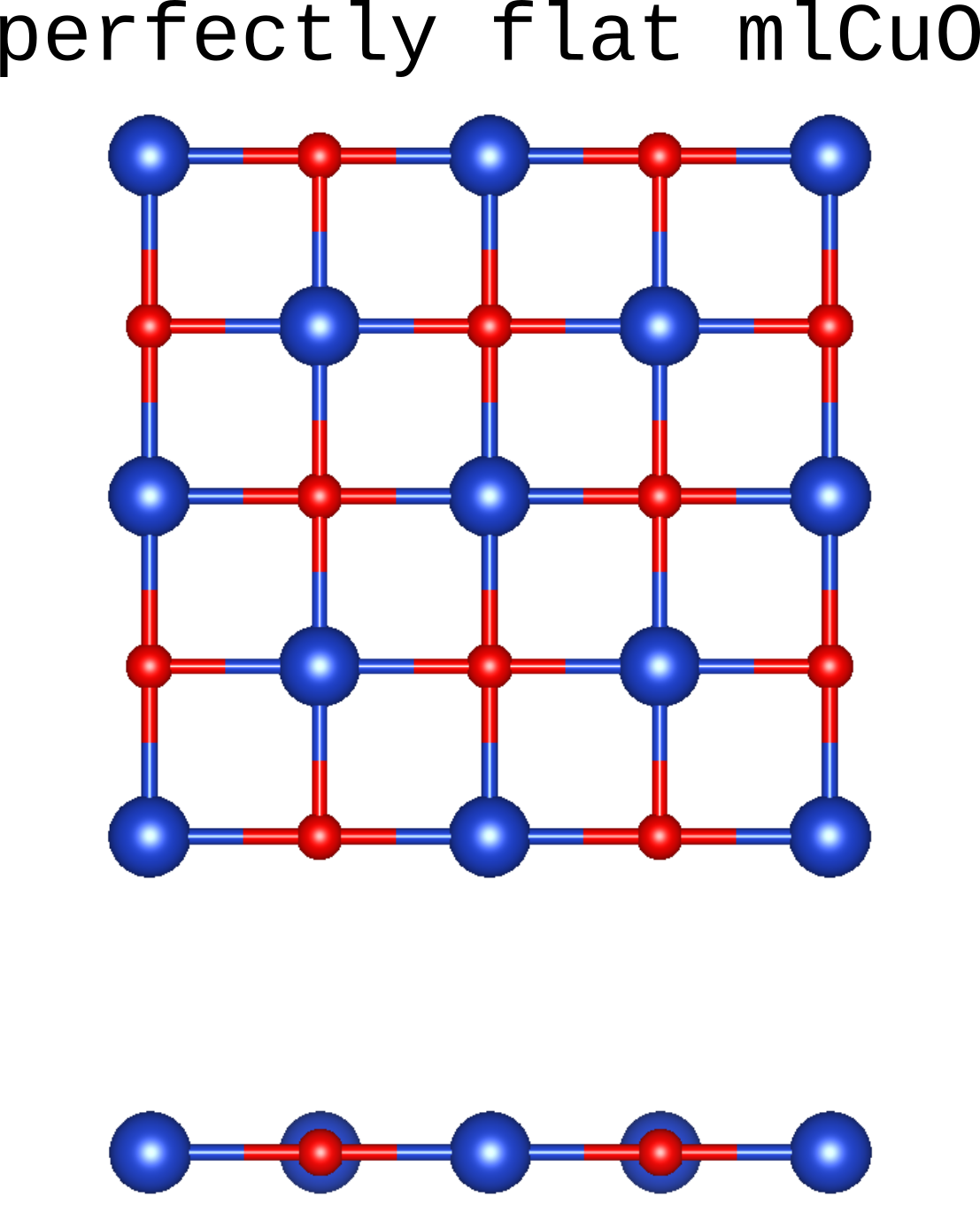}}
\end{minipage}
\hfill
\begin{minipage}[h]{0.23\linewidth}
\center{\includegraphics[width=0.78\linewidth]{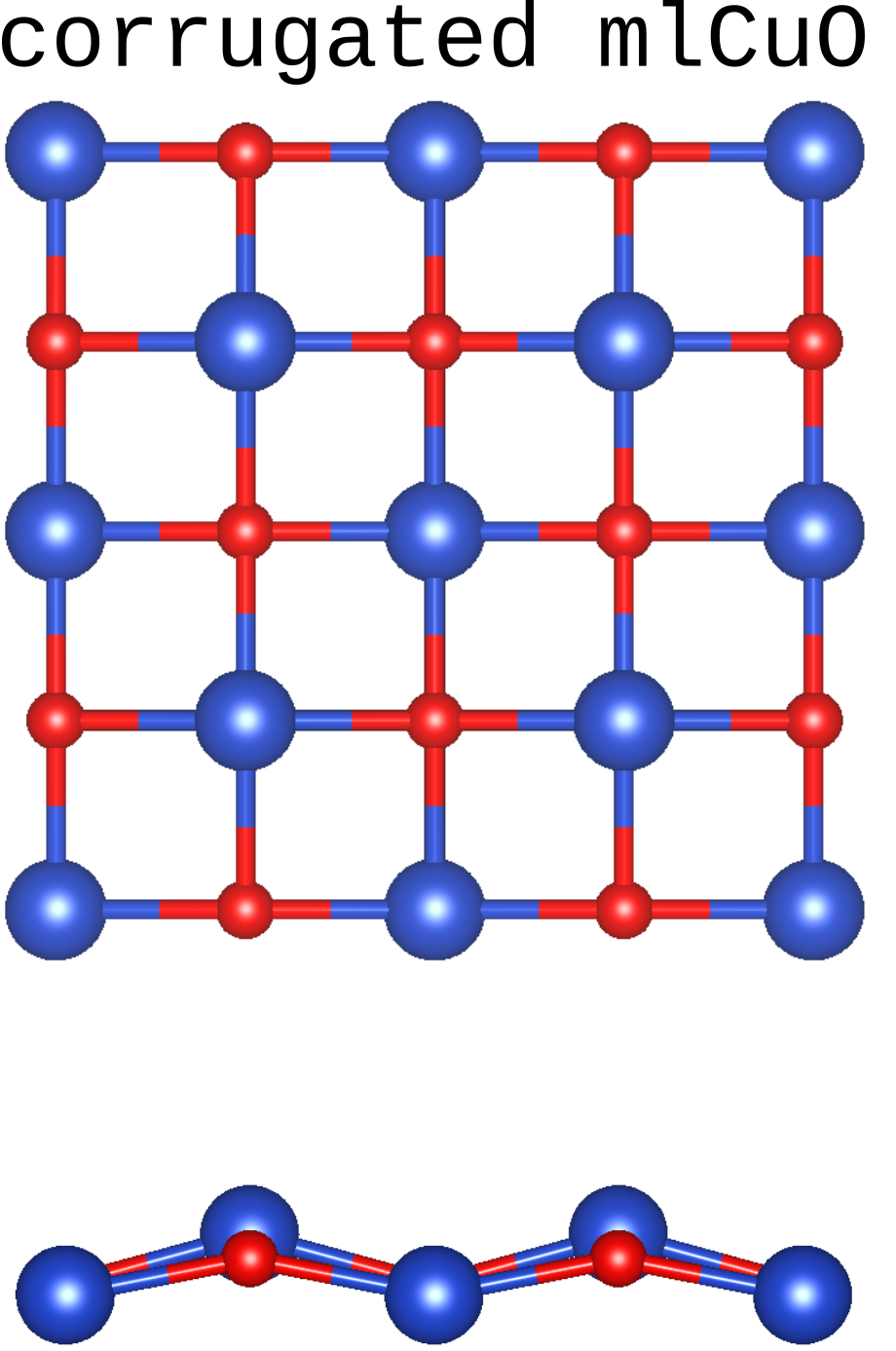}}
\end{minipage}
\hfill
\begin{minipage}[h]{0.23\linewidth}
\center{\includegraphics[width=0.75\linewidth]{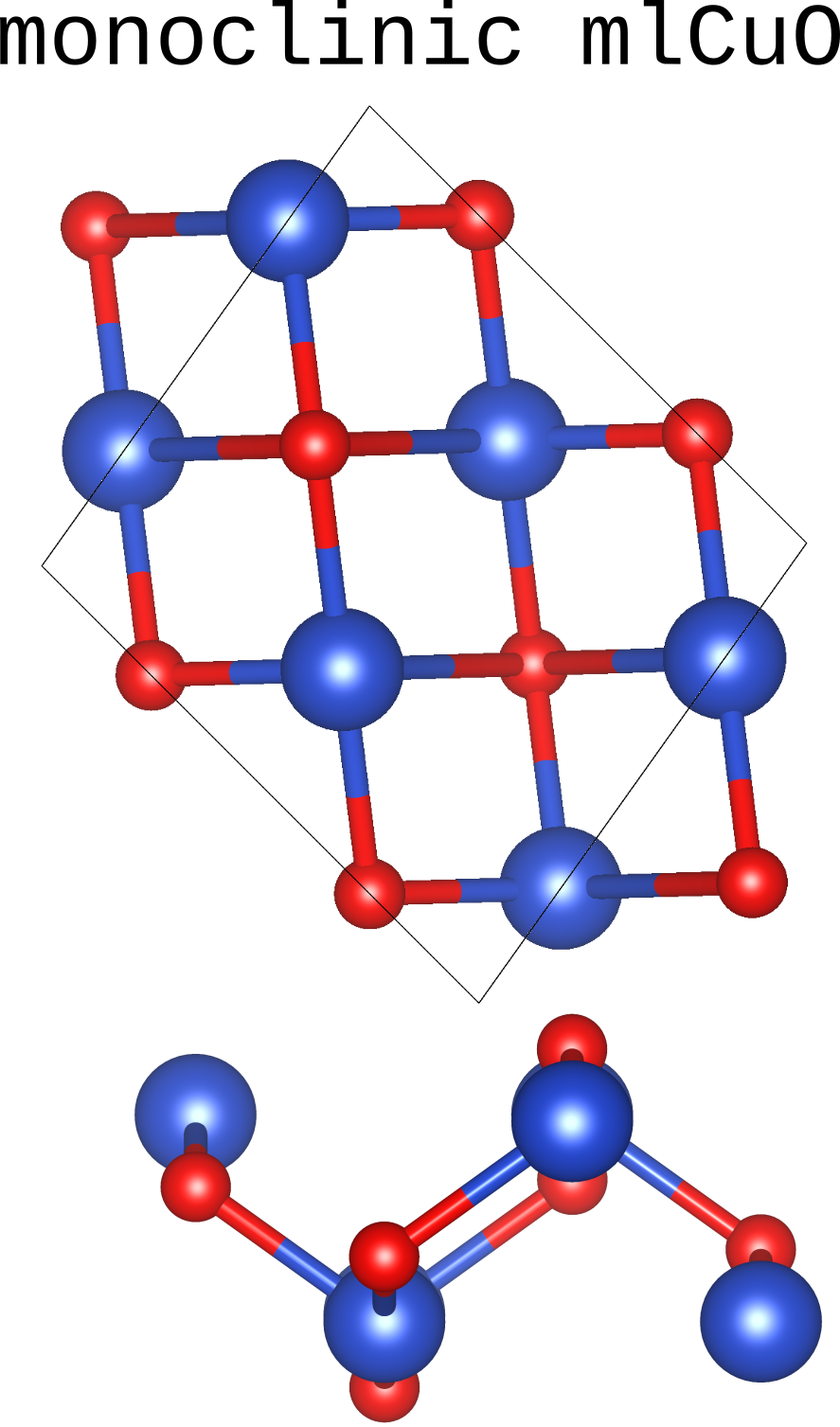}}
\end{minipage}
\begin{minipage}[h]{1\linewidth}
\begin{tabular}{p{0.2\linewidth}p{0.51\linewidth}p{0.23\linewidth}}
\centering (\textbf{a}) & \centering (\textbf{b}) & \centering (\textbf{c})
\end{tabular}
\end{minipage}
\caption{The $\mathrm{mlCuO}$ with the cubic ($2 \times 2 \times 1$ supercell) (\textbf{a}) and monoclinic (\textbf{c}) crystal structure, top and side view (following Ref.~\cite{cao_dft_2018}). Freestanding perfectly flat and corrugated $\mathrm{CuO}$ monolayer crystal structure (\textbf{b}) (following Ref.~\cite{yin_unsupported_2016}). Blue denote $\mathrm{Cu}$ atoms, red is $\mathrm{O}$ atoms. The figures are on the {same spatial scale.}}
\label{ris:structures_bulk}
\end{figure}

Moreover, a number of experiments and theoretical studies were carried out on various structural modifications of the $\mathrm{CuO}$ monolayer: $\mathrm{CuO}$ monolayer in a graphene pores and freestanding $\mathrm{CuO}$ monolayer~\cite{yin_unsupported_2016}, $\mathrm{CuO}$ monolayer on a graphene substrate~\cite{kano_one-atom-thick_2017}, different combinations of $\mathrm{CuO}$ monolayers as an interface between bilayer graphene and finding thermodynamically stable freestanding $\mathrm{CuO}$ monolayer using the evolutionary algorithm~\cite{kvashnin_two-dimensional_2019}.
In Ref.~\cite{yin_unsupported_2016}, the authors showed that freestanding the perfectly flat $\mathrm{CuO}$ monolayer can be corrugated in some cases~(Figure~\ref{ris:structures_bulk}b). This perfectly flat $\mathrm{mlCuO}$ can be easily constructed from the bulk $\mathrm{CuO}$ with the cubic structure.
In Figure~\ref{ris:structures_bulk}a, we show the $2 \times 2 \times 1$ supercell of the cubic $\mathrm{mlCuO}$. It matches with the perfectly flat $\mathrm{mlCuO}$ structure (Figure~\ref{ris:structures_bulk}b, left) when rotated by $45^\circ$. The corrugated $\mathrm{mlCuO}$ can be constructed from the bulk $\mathrm{CuO}$ with the monoclinic structure, though in the monoclinic system atoms are much more displaced; see Figure~\ref{ris:structures_bulk}c and compare it with Figure~\ref{ris:structures_bulk}b, right. Thus, the corrugated state of $\mathrm{mlCuO}$ can be described as a transitional one relative to the $\mathrm{CuO}$ systems with the cubic and the monoclinic structures.

The description of the electronic structure of $\mathrm{mlCuO}$ systems is also far from being complete---the authors were able to find some data in Ref.~\cite{yin_unsupported_2016,kvashnin_two-dimensional_2019}, but no more than that. Besides, at the moment the authors are not aware of any works where a minimal model has been formulated neither for an isolated perfectly flat monolayer $\mathrm{CuO}$, nor for more complex crystal structures such as the corrugated monolayer, monolayer on a substrate or monolayer as an interface. Thus, it seems necessary to obtain on a more systematic basis the densities of states, the band structures and the Fermi surfaces for all listed monolayer $\mathrm{CuO}$ systems and to formulate a minimal model for them with the corresponding Hamiltonian \mbox{parameter values.}

In this work, we solve a task of proposing and comparing different minimal models for the $\mathrm{CuO}$ monolayer systems as a necessary first step of any further theoretical investigations.


\section{Crystal Structure and Calculation Details}

To calculate the band structure, the density of states (DOS) and the Fermi surface, we used the density functional theory with the full-potential linear augmented plane-wave framework, as implemented in \texttt{{WIEN2k}}~\cite{wien2k} together with the generalized gradient approximation by Perdew, Burke and Ernzerhof~\cite{pbe}, to the exchange-correlation functional.

\begin{figure}[H]
\begin{minipage}[h]{0.2\linewidth}
\center{\includegraphics[width=0.7\linewidth]{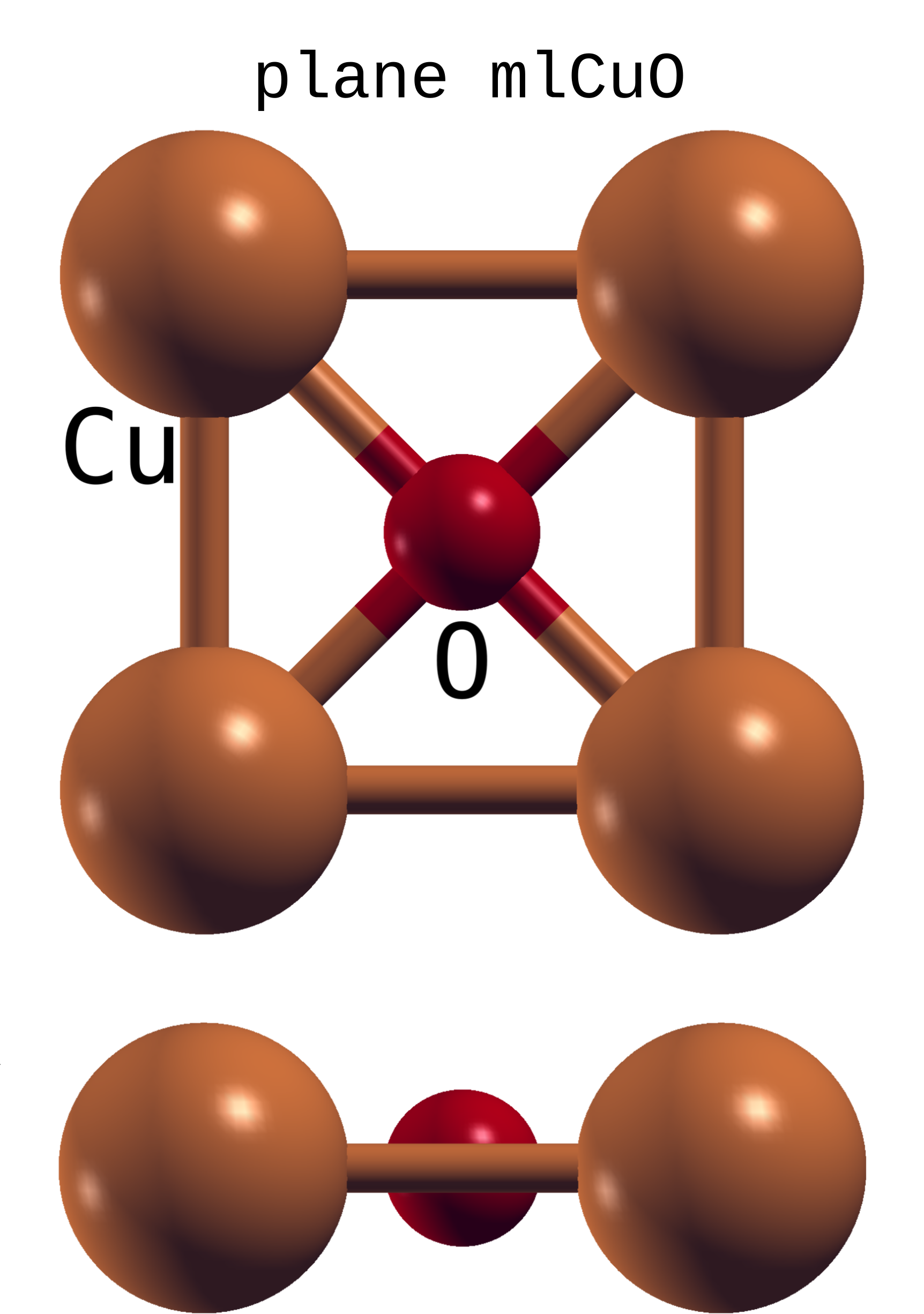}}
\end{minipage}
\begin{minipage}[h]{0.2\linewidth}
\center{\includegraphics[width=0.7\linewidth]{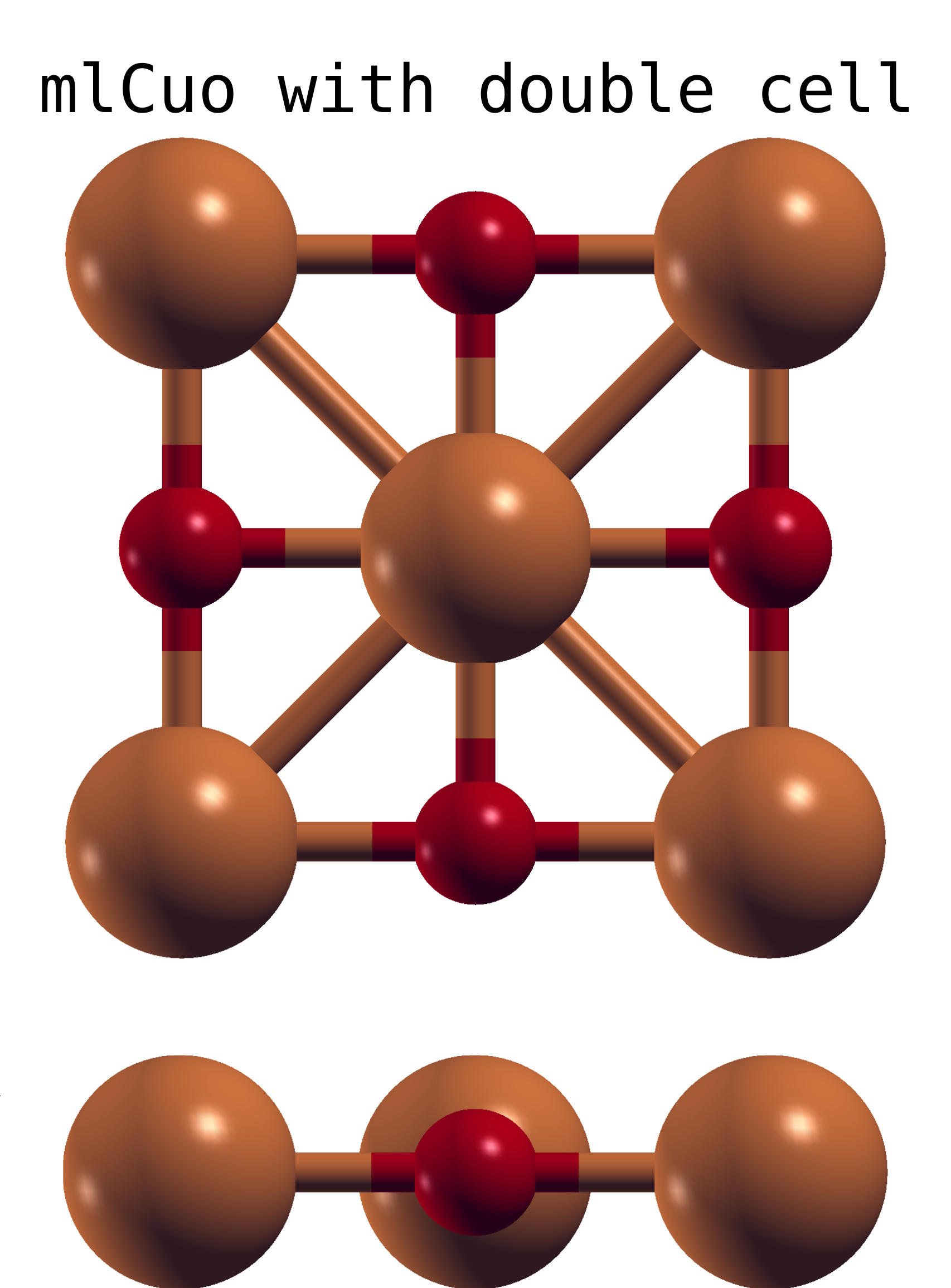}}
\end{minipage}
\begin{minipage}[h]{0.2\linewidth}
\center{\includegraphics[width=0.7\linewidth]{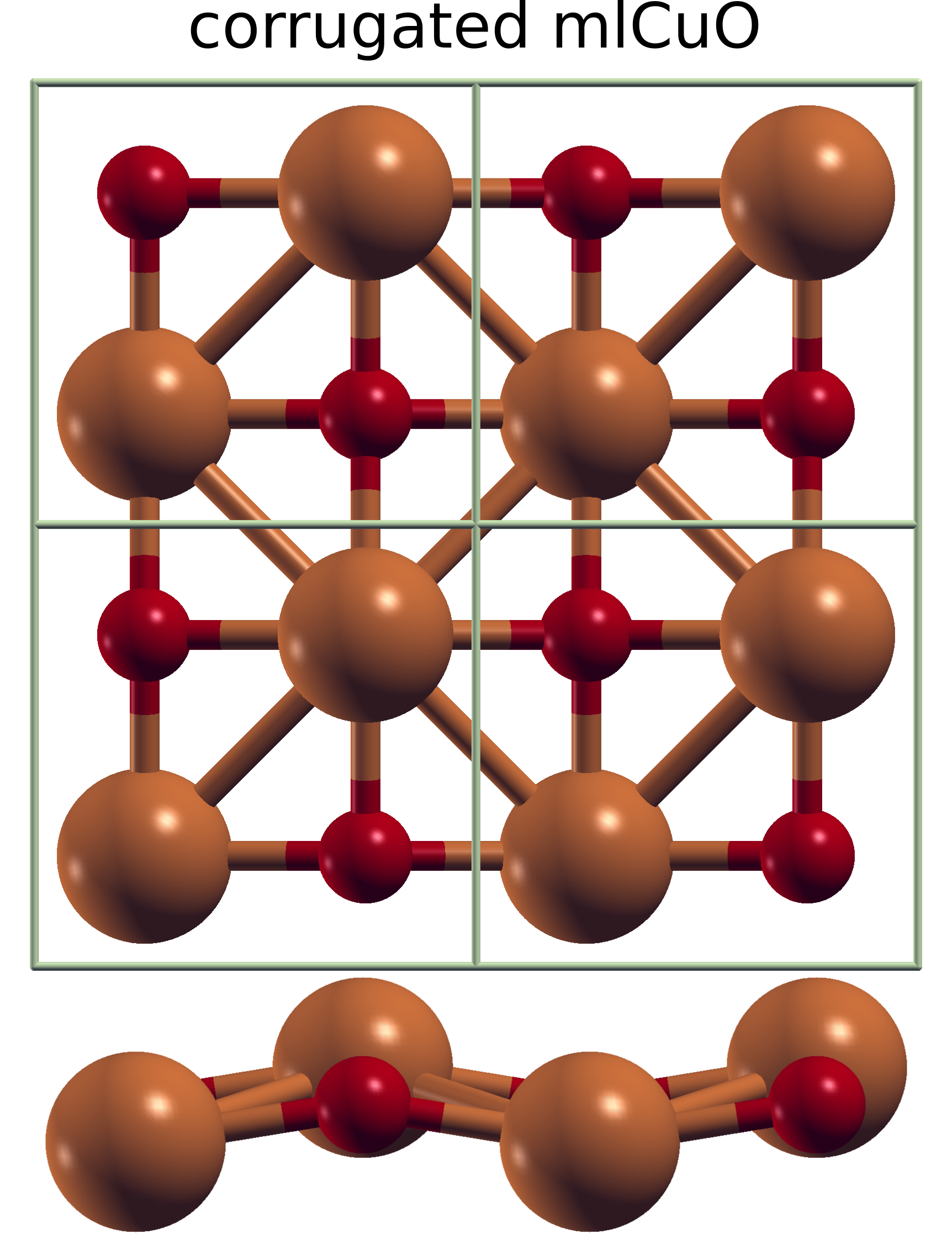}}
\end{minipage}
\begin{minipage}[h]{0.2\linewidth}
\center{\includegraphics[width=0.95\linewidth]{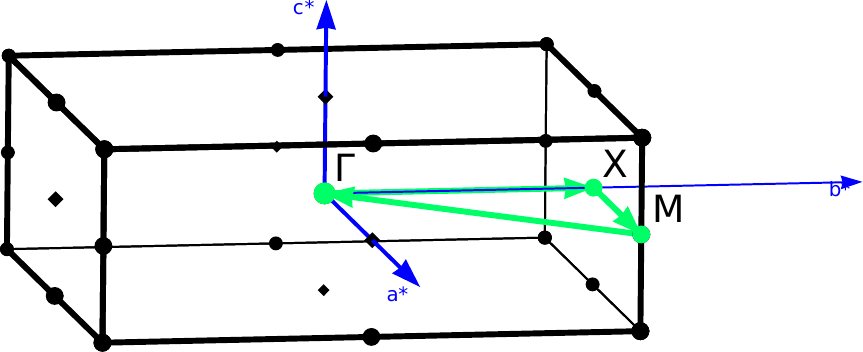}}
\end{minipage}
\begin{minipage}[h]{1\linewidth}
\begin{tabular}{p{0.17\linewidth}p{0.305\linewidth}p{0.17\linewidth}p{0.26\linewidth}}
\centering (\textbf{a}) & \centering (\textbf{b}) & \centering (\textbf{c}) & \centering (\textbf{d})
\end{tabular}
\end{minipage}
\caption{The flat $\mathrm{mlCuO}$ (\textbf{a}), the $\mathrm{mlCuO}$ with a doubled unit cell (\textbf{b}), the corrugated $\mathrm{mlCuO}$ (\mbox{$2 \times 2 \times 1$ supercell}) (\textbf{c}) crystal structures considered in this work, top and side view; brown denote $\mathrm{Cu}$ atoms, red is $\mathrm{O}$ atoms. Brillouin zone for the doubled unit cell and corrugated $\mathrm{mlCuO}$~(\textbf{d}).}
\label{ris:structures}
\end{figure}

Figure~\ref{ris:structures} shows the crystal structures of the systems discussed in this paper. The flat $\mathrm{CuO}$ monolayer space group is a 123 (\textit{P4/mmm}). The lattice parameter is $a = 2.69$~\AA~\cite{yin_unsupported_2016}. Atoms occupy the following positions: $\mathrm{Cu}$ 1a $(0,0,0)$ and $\mathrm{O}$ 1c $(0.5,0.5,0)$. We used a 20~Bohr vacuum gap. In order to construct the corrugated $\mathrm{CuO}$ monolayer, we doubled the unit cell and made these new additional $\mathrm{Cu}$~and~$\mathrm{O}$ atoms unequivalent to the original ones by applying a small shift about 0.5~Bohr in \textit{z} direction only for them. Resulting system has 59 (\textit{Pmmn}) [origin choice 2] space group. Next, we did a set of structural relaxations with 10, 20 and 40~Bohr vacuum gaps. There was no difference between 20 and 40~Bohr vacuum gap cases, so in all further calculations we used the 20~Bohr gap. After relaxation neighboring $\mathrm{Cu}$ atoms shifted in \textit{z} direction about $\pm 0.31$~\AA~relative to their original positions, while neighboring $\mathrm{O}$ atoms barely shifted at all. Since the final corrugated structure of $\mathrm{CuO}$ monolayer has a doubled unit cell and is rotated by $45^\circ$ relative to the flat $\mathrm{mlCuO}$, we cannot directly compare their calculated electronic structures. Thus, we use additional system---the flat $\mathrm{mlCuO}$ with a doubled (and rotated) unit cell---in order to make a proper comparison. \xreplaced{Moreover, for the flat $\mathrm{mlCuO}$ we did a $45^\circ$ rotation of a local coordinate system in order to use a typical orbital convention, such as in cuprate compounds.}{Moreover, for both of these systems, we did a $45^\circ$ rotation of a local coordinate system.}

All calculations were nonmagnetic and converged self-consistently on a grid of \mbox{24 × 24 × 1} \textit{k}-points in the irreducible Brillouin zone using the Monkhorst-Pack method~\cite{monkhorst1976special}. \xadded{We used energy convergence limit 0.1~mRy, force convergence limit 0.5~mRy/a.u. for optimization, $\text{RKmax} = 7$, $\text{Gmax} = 12$, energy separation $-6.0$~Ry.} In Figure~\ref{ris:structures}d, we show the Brillouin zone with the \textit{k}-path used in the band structure analysis.

\section{Results and Discussion}

\subsection{Electronic Structure}

\begin{figure}[H]
\begin{minipage}[h]{0.38\linewidth}
\center{\includegraphics[width=1\linewidth]{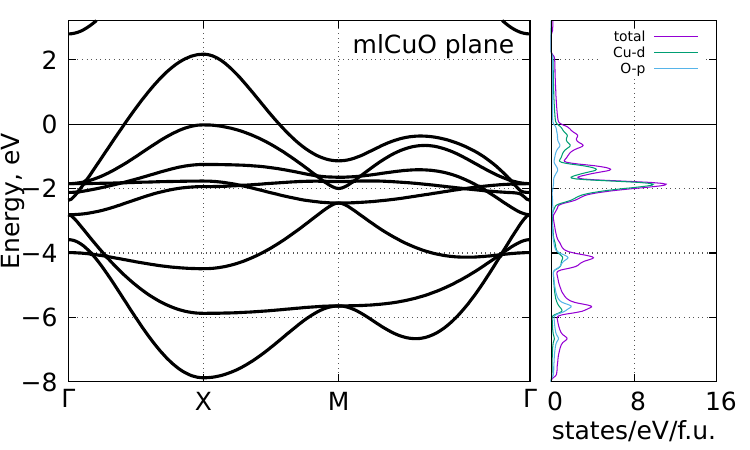}}
\end{minipage}
\hfill
\begin{minipage}[h]{0.37\linewidth}
\center{\includegraphics[width=1\linewidth]{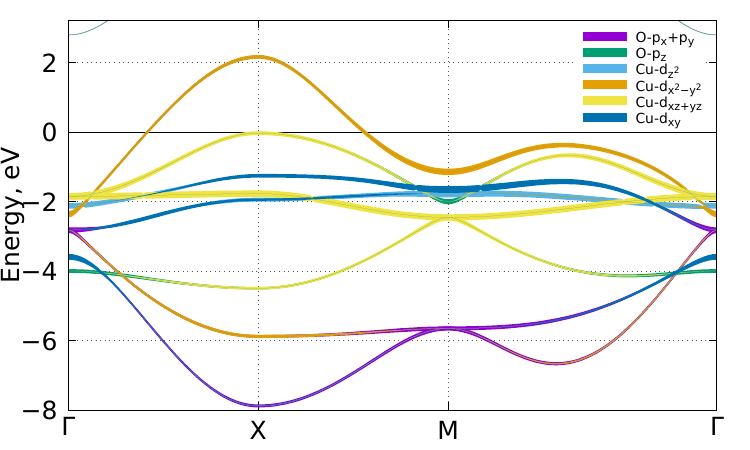}}
\end{minipage}
\hfill
\begin{minipage}[h]{0.21\linewidth}
\center{\includegraphics[width=1\linewidth]{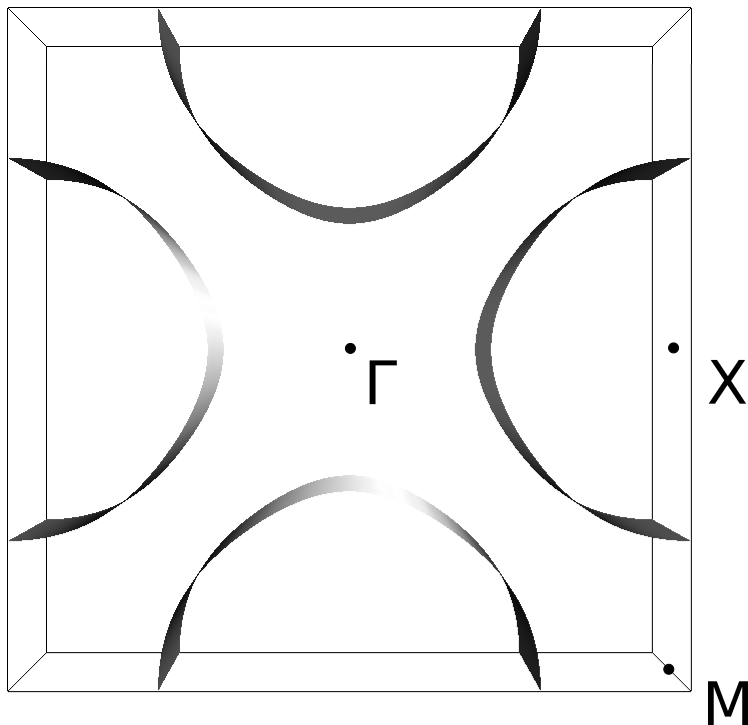}}
\end{minipage}
\begin{minipage}[h]{1\linewidth}
\begin{tabular}{p{0.33\linewidth}p{0.43\linewidth}p{0.12\linewidth}}
\centering (\textbf{a}) & \centering (\textbf{b}) & \centering (\textbf{c}) \\
\end{tabular}
\end{minipage}
\vfill
\begin{minipage}[h]{0.38\linewidth}
\center{\includegraphics[width=1\linewidth]{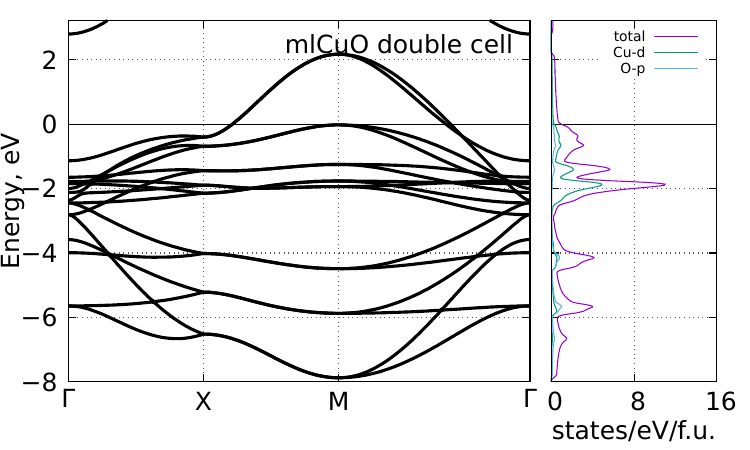}}
\end{minipage}
\hfill
\begin{minipage}[h]{0.37\linewidth}
\center{\includegraphics[width=1\linewidth]{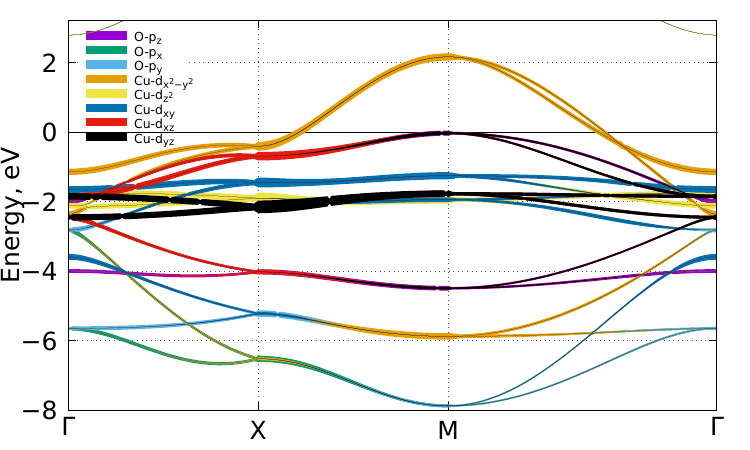}}
\end{minipage}
\hfill
\begin{minipage}[h]{0.21\linewidth}
\center{\includegraphics[width=1\linewidth]{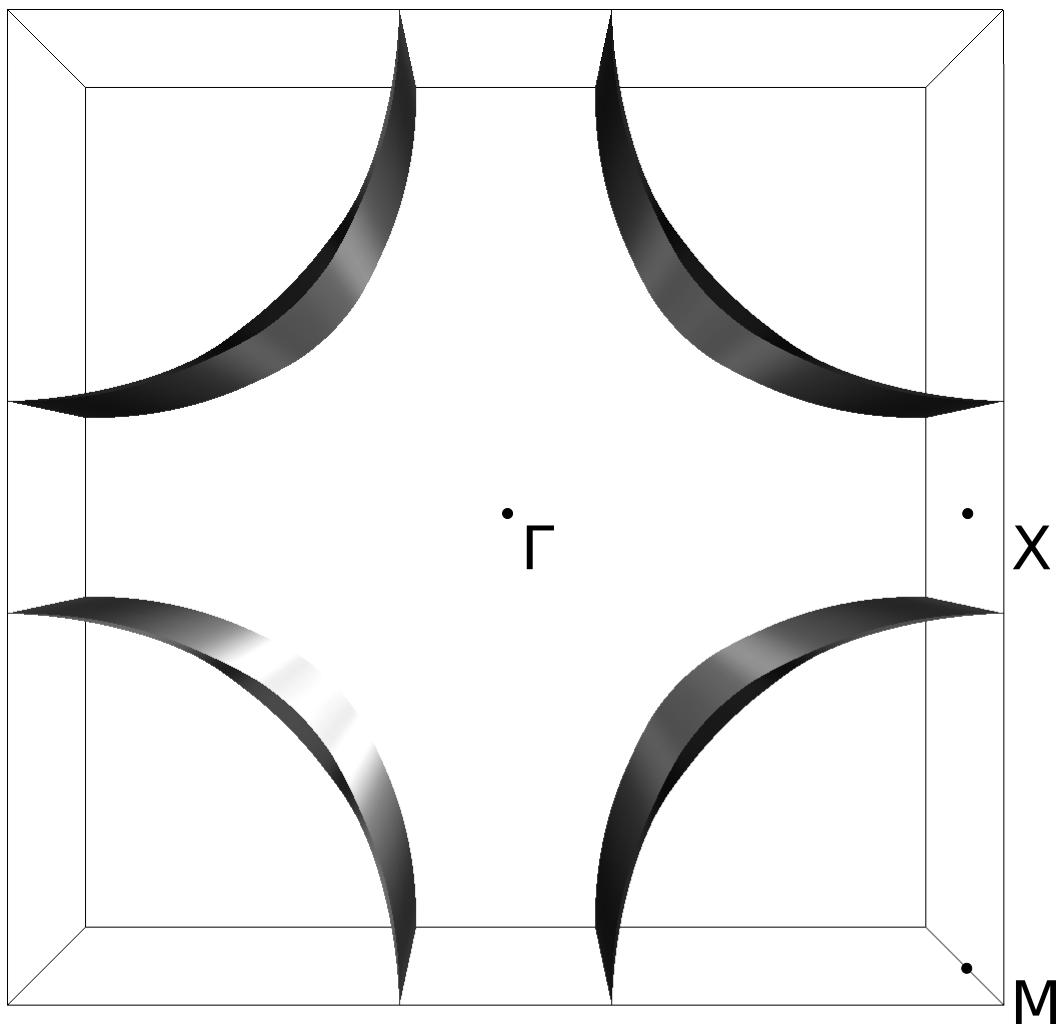}}
\end{minipage}
\begin{minipage}[h]{1\linewidth}
\begin{tabular}{p{0.33\linewidth}p{0.43\linewidth}p{0.12\linewidth}}
\centering (\textbf{d}) & \centering (\textbf{e}) & \centering (\textbf{f}) \\
\end{tabular}
\end{minipage}
\vfill
\begin{minipage}[h]{0.38\linewidth}
\center{\includegraphics[width=1\linewidth]{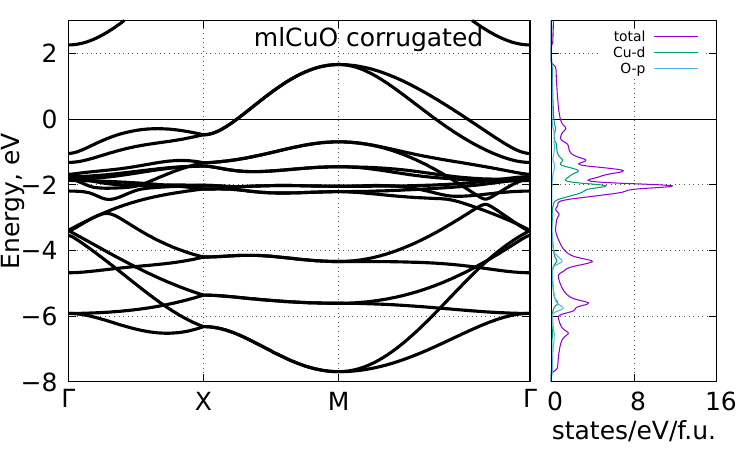}}
\end{minipage}
\hfill
\begin{minipage}[h]{0.37\linewidth}
\center{\includegraphics[width=1\linewidth]{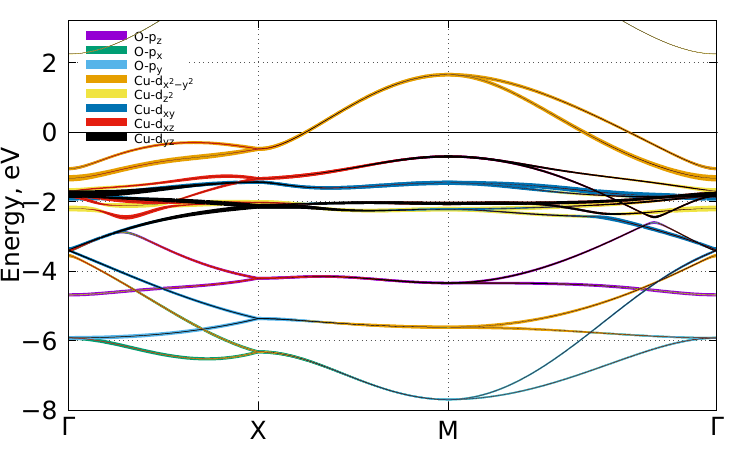}}
\end{minipage}
\hfill
\begin{minipage}[h]{0.21\linewidth}
\center{\includegraphics[width=1\linewidth]{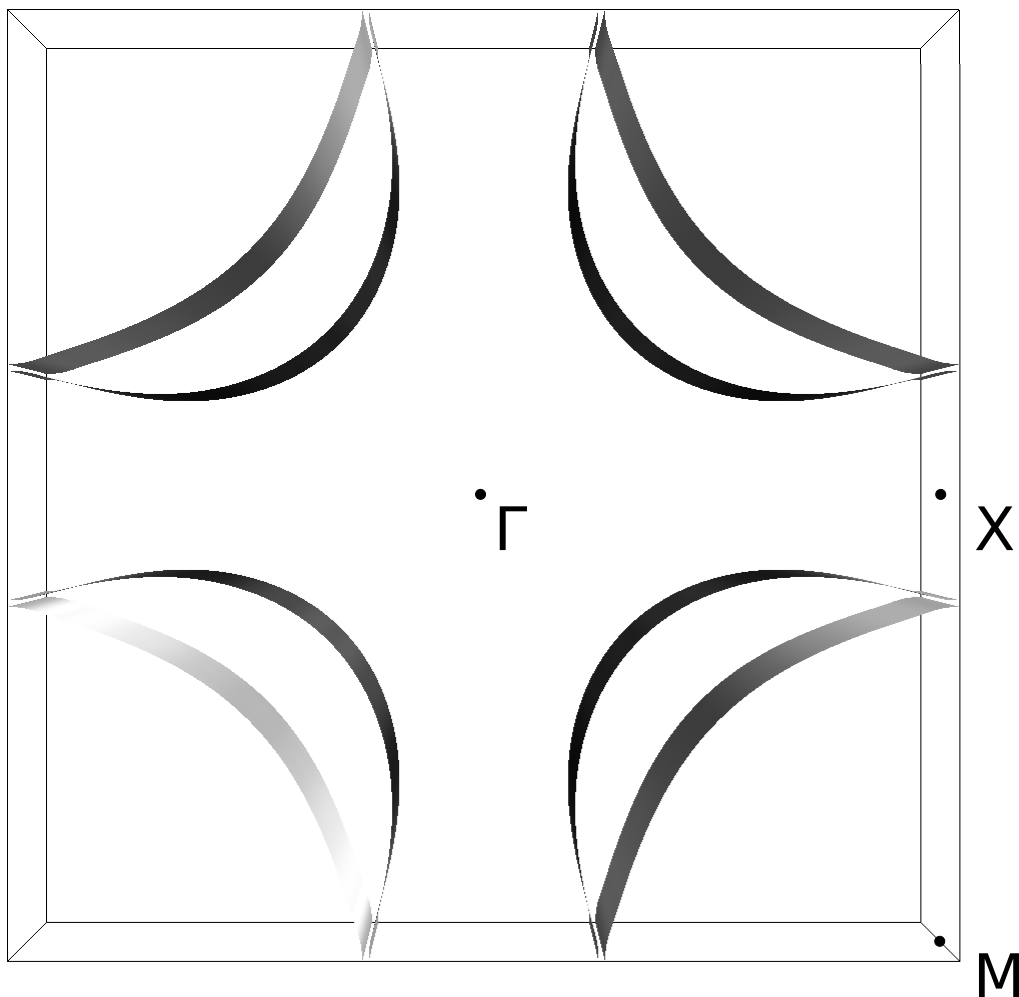}}
\end{minipage}
\begin{minipage}[h]{1\linewidth}
\begin{tabular}{p{0.33\linewidth}p{0.43\linewidth}p{0.12\linewidth}}
\centering (\textbf{g}) & \centering (\textbf{h}) & \centering (\textbf{i})
\end{tabular}
\end{minipage}
\caption{DFT (GGA) calculated DOS, the band structures, the band structures with the orbital characters and the Fermi surface of the flat $\mathrm{mlCuO}$ (\textbf{a}--\textbf{c}), the $\mathrm{mlCuO}$ with a doubled unit cell (\textbf{d}--\textbf{f}), the corrugated $\mathrm{mlCuO}$ (\textbf{g}--\textbf{i}). Zero corresponds to the Fermi level.}
\label{ris:bands_dos_fs}
\end{figure}

Figure~\ref{ris:bands_dos_fs} shows the DFT (GGA) band structures, the densities of states, bands with their orbital characters and the Fermi surfaces. The first row~(a--c) of Figure~\ref{ris:bands_dos_fs} shows the results for the flat $\mathrm{mlCuO}$. The band structure in Figure~\ref{ris:bands_dos_fs}a shows that there is an isolated set of bands in the range from $-8$~eV to $2.3$~eV resembling typical $\mathrm{Cu}$-based HTSC $\mathrm{La_2CuO_4}$. It has only the $\mathrm{Cu}$-$3d$ and the $\mathrm{O}$-$2p$ states. The electronic bands of the flat $\mathrm{mlCuO}$ at the Fermi level are formed by the $\mathrm{Cu}$-$3d_{x^2-y^2}$ states (with small addition of the hybrid $\mathrm{O}$-$2p$ states) in consistence with the known results~\cite{ekuma_electronic_2014}. Note that there is a second band that almost crosses the Fermi level---it is only 0.02~eV lower. It includes the $\mathrm{Cu}$-$3d_{xz,yz}$ states hybridized with the $\mathrm{O}$-$2p_z$. The Fermi surface has a hole pocket around the $X$ point.

The second row~(d--f) of Figure~\ref{ris:bands_dos_fs} shows the results for the $\mathrm{mlCuO}$ with a doubled unit cell. Due to $45^\circ$ rotation of a unit cell, its band structure differs from that of the flat $\mathrm{mlCuO}$ in a much more complex way than simply by a large number of bands. The two bands crossing the Fermi level originate from two $\mathrm{Cu}$ atoms in the unit cell. They are formed by the \xadded{$\mathrm{Cu}$-$3d_{x^2-y^2}$} states. Regarding them, we can note degeneracy lifting in the $\Gamma - M$ direction. As in the flat $\mathrm{mlCuO}$, there are the $\mathrm{Cu}$-$3d_{xz,yz}$ states (four bands) just below the Fermi level. The Fermi surface has two hole pockets around the $M$ point.

The third row~(g--i) of Figure~\ref{ris:bands_dos_fs} shows the results for the corrugated $\mathrm{mlCuO}$. Its electronic structure is rather similar to that in the $\mathrm{mlCuO}$ with a doubled unit cell, but there are two notable differences. First, a significant shift of the $\mathrm{Cu}$-$3d_{xz,yz}$ bands to $-0.7$~eV. Second, a noticeably larger degeneracy lifting for the \xadded{$\mathrm{Cu}$-$3d_{x^2-y^2}$} bands in the $\Gamma - M$ direction. On top of that, the corrugated $\mathrm{mlCuO}$ total energy turns out to be lower than the flat $\mathrm{mlCuO}$ one by $0.07$~eV. In other words, the corrugated state appears to be more favorable and if we have the flat $\mathrm{mlCuO}$ as a topmost layer of some surface, it will most likely be corrugated.

Seeing such a significant shift of the $\mathrm{Cu}$-$3d_{xz,yz}$ bands of the corrugated $\mathrm{mlCuO}$, we wondered if it was possible to raise these states to the Fermi level only via the lattice strain.
Significant lattice strain is observed at all sorts of interfaces where there is a mismatch between the lattice parameters.
Besides, as we already mention, the corrugated state appears to be more energetically favorable.

\begin{figure}[H]
\begin{minipage}[H]{0.165\linewidth}
\center{\includegraphics[width=1\linewidth]{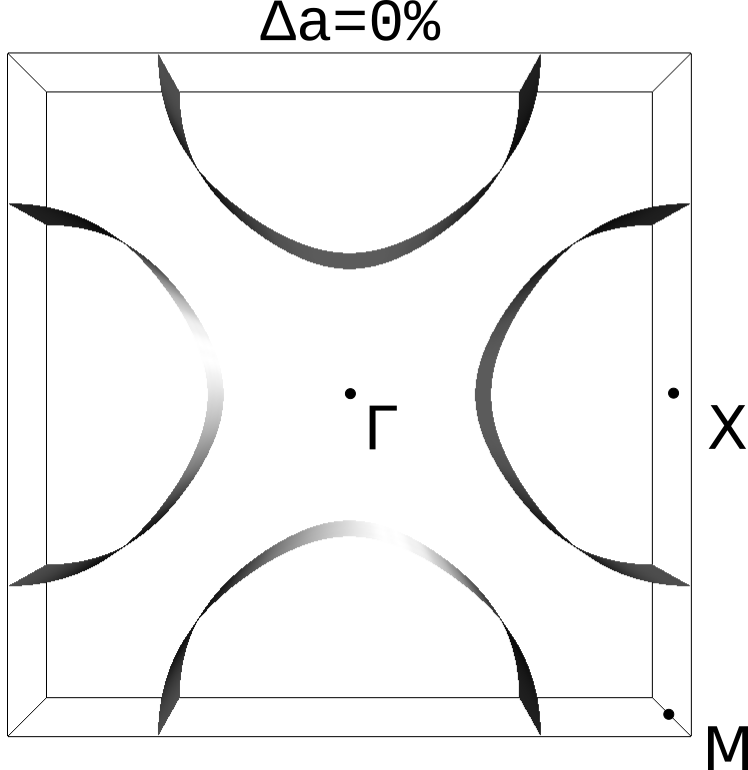}}
\end{minipage}
\hfill
\begin{minipage}[h]{0.165\linewidth}
\center{\includegraphics[width=1\linewidth]{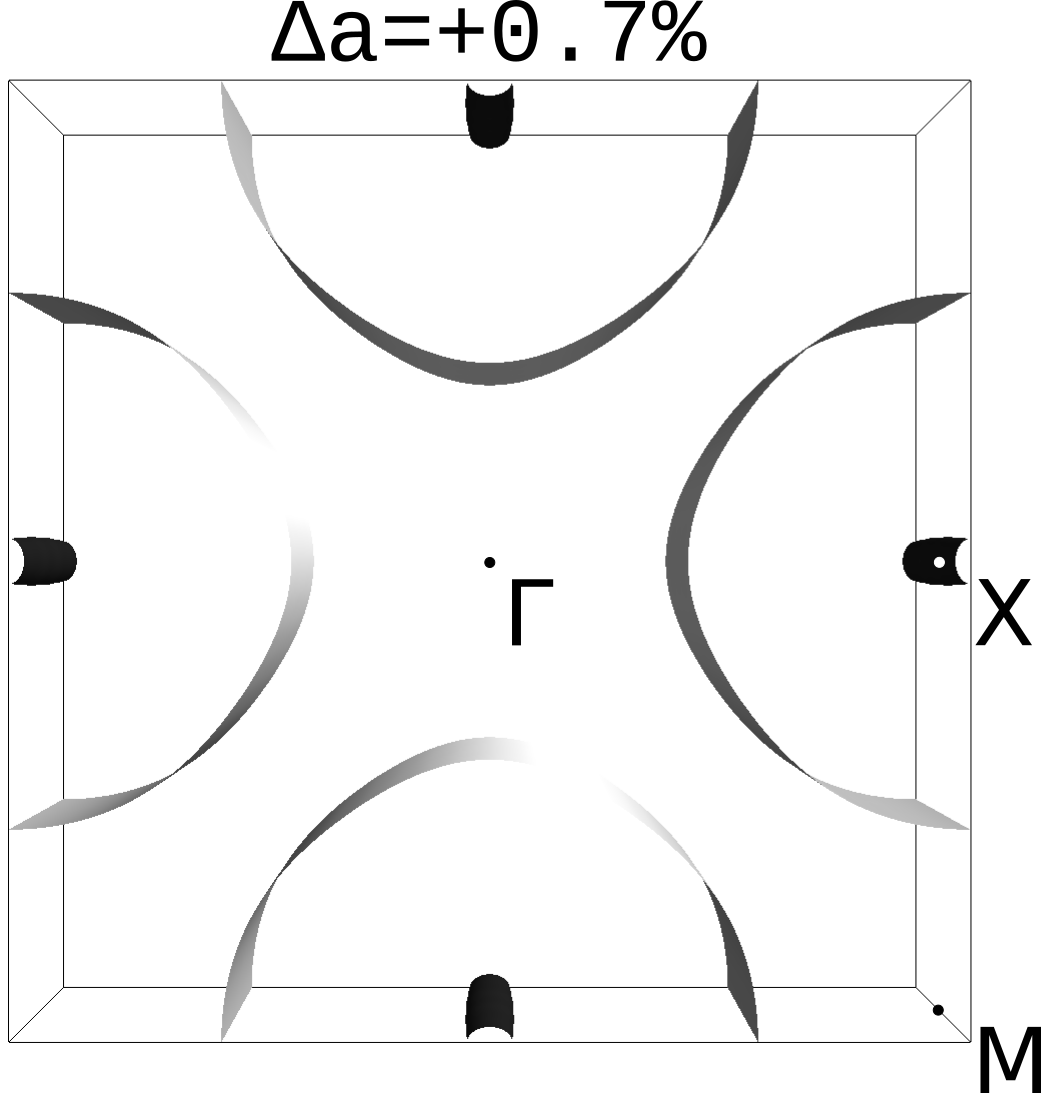}}
\end{minipage}
\hfill
\begin{minipage}[h]{0.165\linewidth}
\center{\includegraphics[width=1\linewidth]{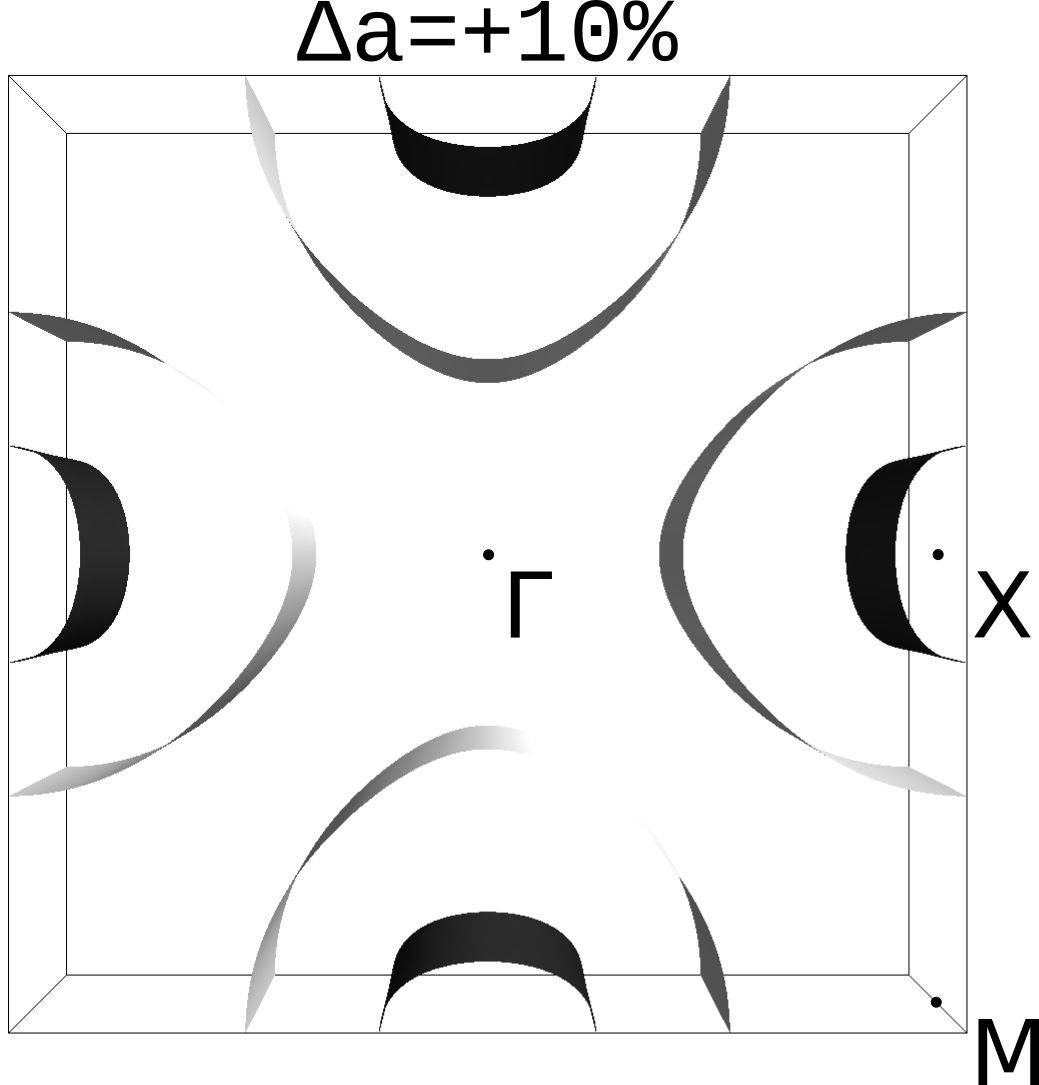}}
\end{minipage}
\hfill
\begin{minipage}[h]{0.43\linewidth}
\center{\includegraphics[width=1\linewidth]{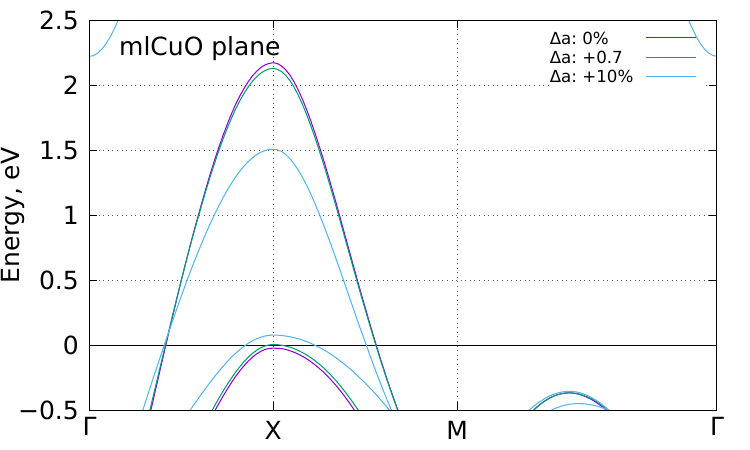}}
\end{minipage}
\begin{minipage}[h]{1\linewidth}
\begin{tabular}{p{0.14\linewidth}p{0.17\linewidth}p{0.15\linewidth}p{0.46\linewidth}}
\centering (\textbf{a}) & \centering (\textbf{b}) & \centering (\textbf{c}) & \centering (\textbf{d}) \\
\end{tabular}
\end{minipage}
\vfill
\begin{minipage}[h]{0.24\linewidth}
\center{\includegraphics[width=1\linewidth]{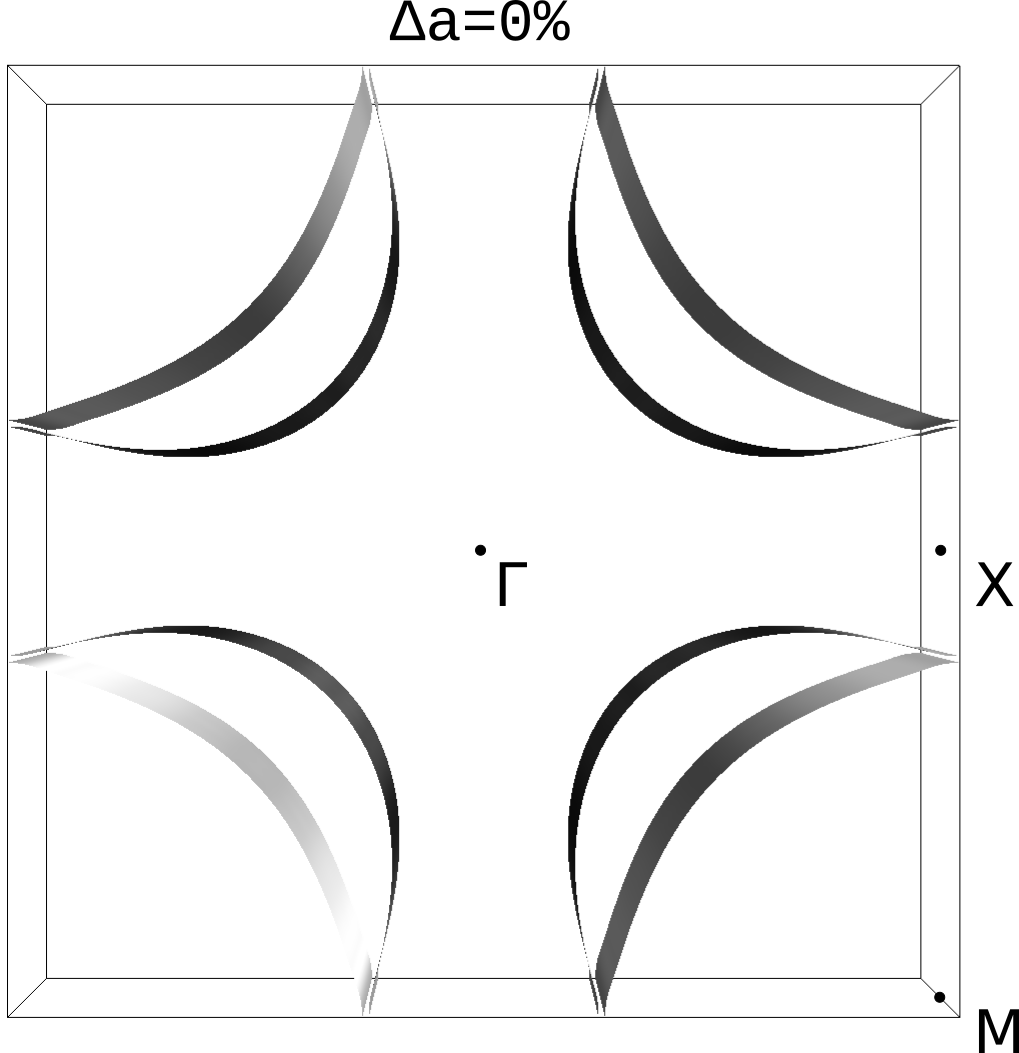}}
\end{minipage}
\hfill
\begin{minipage}[h]{0.24\linewidth}
\center{\includegraphics[width=1\linewidth]{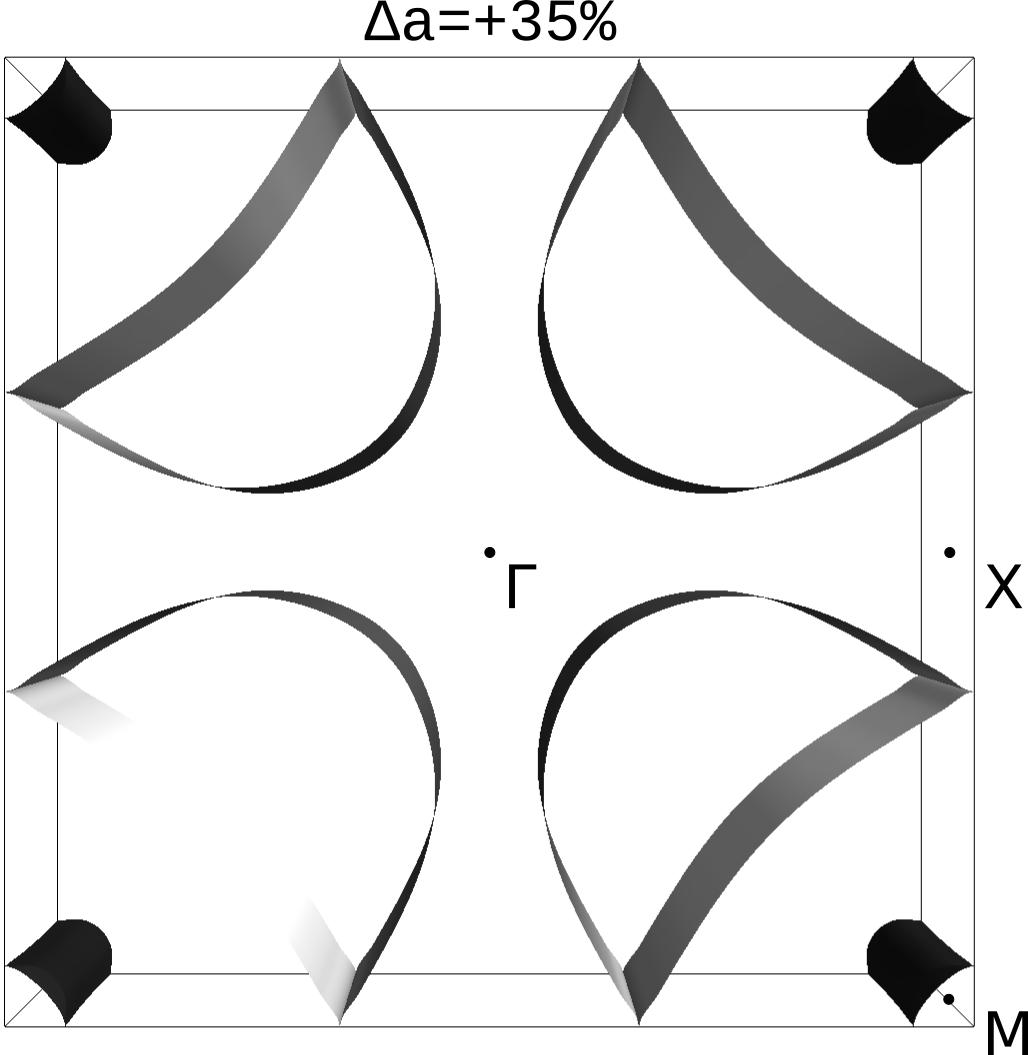}}
\end{minipage}
\hfill
\begin{minipage}[h]{0.43\linewidth}
\center{\includegraphics[width=1\linewidth]{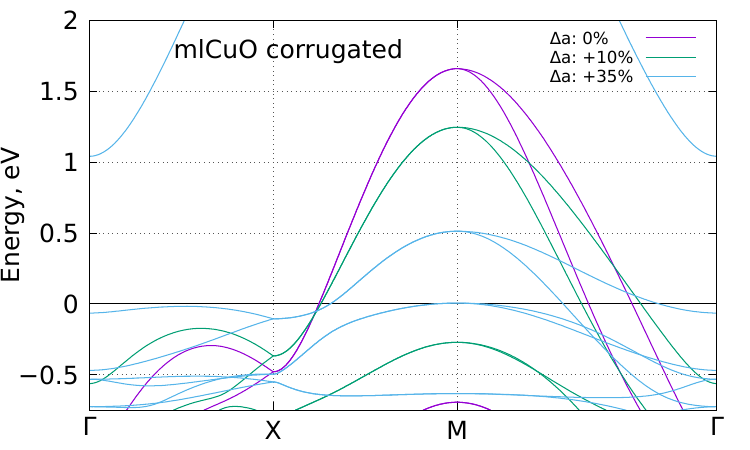}}
\end{minipage}
\begin{minipage}[h]{1\linewidth}
\begin{tabular}{p{0.21\linewidth}p{0.29\linewidth}p{0.45\linewidth}}
\centering (\textbf{e}) & \centering (\textbf{f}) & \centering (\textbf{g})
\end{tabular}
\end{minipage}
\caption{DFT (GGA) calculated Fermi surface with $\Delta a = 0\%$ (zero strain), $\Delta a = +0.7\%$, $\Delta a = +10\%$ and the band structure comparison for the flat $\mathrm{mlCuO}$ (\textbf{a}--\textbf{d}); the Fermi surface with $\Delta a = 0\%$ (zero strain), $\Delta a = +35\%$ and the band structure comparison for the corrugated $\mathrm{mlCuO}$ (\textbf{e}--\textbf{g}). Zero corresponds to the Fermi level.}
\label{ris:strain_fs_bands}
\end{figure}

To clarify this issue, we carried out a series of calculations where we varied the lattice parameter $a$ from 0\% to 10\% for the flat $\mathrm{mlCuO}$ and from 0\% to 35\% for the corrugated $\mathrm{mlCuO}$. The corresponding results can be observed in Figure~\ref{ris:strain_fs_bands}. For the flat $\mathrm{mlCuO}$, the lattice deformation $\Delta a = 0.7\%$ leads to a topological Lifshitz transition with the appearance of a new hole pocket around the $X$ point. Clearly, it is a very minor lattice parameter change of the order of experiment accuracy. For the corrugated $\mathrm{CuO}$ monolayer, such a transition requires a much larger lattice deformation; it appears only at $\Delta a = 35\%$. Of course, such a strain is too large, and we bring it here only as an illustration. However, for the corrugated $\mathrm{mlCuO}$ case, we want to note a presence of what seems to be a flat band in the $\Gamma - X$ direction near the Fermi level. It is likely that a flat band at the Fermi level can be obtained using a reasonable lattice strain and a hole doping.

Let us now return to the idea we proposed in the introduction: the corrugated of $\mathrm{mlCuO}$ can be described as a transitional system in between the $\mathrm{CuO}$ with the cubic and monoclinic structures~(Figure~\ref{ris:structures_bulk}). Table~\ref{table_dist} shows $\mathrm{Cu}$-$\mathrm{Cu}$ and $\mathrm{Cu}$-$\mathrm{O}$ distances in the bulk $\mathrm{CuO}$ systems and the $\mathrm{CuO}$ monolayers.
In the bulk $\mathrm{CuO}$ with the cubic structure, $\mathrm{CuO}$ layers are significantly stretched---by about $11.5\%$ relative to the flat $\mathrm{mlCuO}$, which is close to the considered lattice strain $\Delta a = 10\%$ (Figure~\ref{ris:strain_fs_bands}c). We studied the monolayer made from the bulk $\mathrm{CuO}$ by simply adding a vacuum between layers (getting slab+vacuum) in \textit{z} direction and using no relaxation~(Figure~\ref{ris:structures_bulk}a). The final crystal and band structures were nearly identical to the flat $\mathrm{mlCuO}$ ones with lattice strain $\Delta a = 10\%$ (as in Figure~\ref{ris:strain_fs_bands}c,d); thus, we do not include them.

\begin{table}[H]
\caption{$\mathrm{Cu}$-$\mathrm{Cu}$ and $\mathrm{Cu}$-$\mathrm{O}$ distances in the bulk and the monolayer $\mathrm{CuO}$ systems.}
\label{table_dist}

 \newcolumntype{M}[1]{>{\raggedright\arraybackslash}m{#1}}
\begin{tabularx}{\textwidth}{p{4cm}p{2.5cm}p{2.5cm}p{2.5cm}}
\toprule
\textbf{System} & \textbf{d($\mathrm{Cu}$-$\mathrm{Cu}$), \AA} & \textbf{d($\mathrm{Cu}$-$\mathrm{O}$), \AA} & \textbf{Ref.} \\
\midrule
Cubic bulk $\mathrm{CuO}$ & $3.00$ & $2.12$ & \cite{schmahl_uber_1968} \\
Monoclinic bulk $\mathrm{CuO}$ & $2.90$ &	 $1.96$ & \cite{asbrink_refinement_1970} \\
Flat $\mathrm{mlCuO}$ & $2.69$ & $1.90$ & This work,~ref \cite{yin_unsupported_2016} \\
Corrugated $\mathrm{mlCuO}$ & $2.76$ & $1.93$ & This work \\
\bottomrule
\end{tabularx}

\end{table}

\begin{figure}[H]
\begin{minipage}[h]{0.32\linewidth}
\center{\includegraphics[width=1\linewidth]{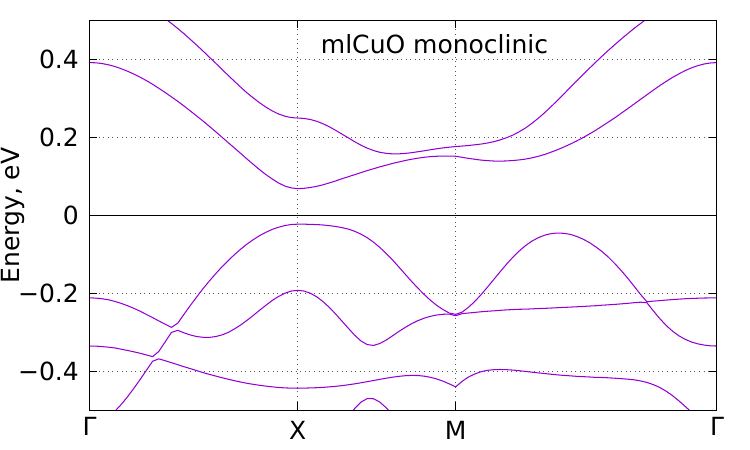}}
\end{minipage}
\hfill
\begin{minipage}[h]{0.32\linewidth}
\center{\includegraphics[width=1\linewidth]{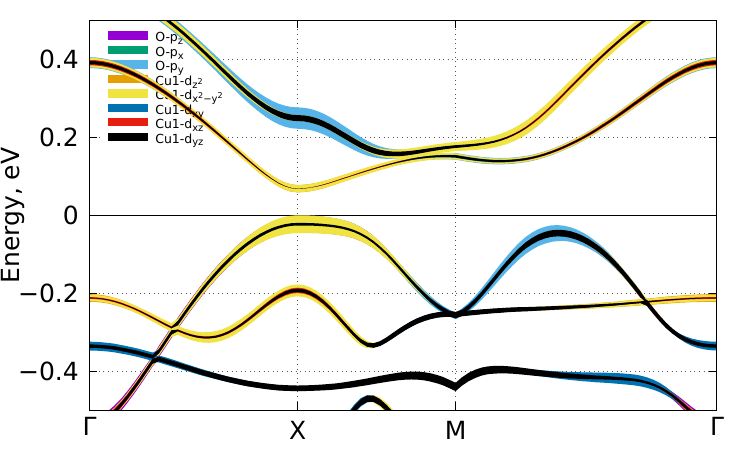}}
\end{minipage}
\hfill
\begin{minipage}[h]{0.32\linewidth}
\center{\includegraphics[width=0.95\linewidth]{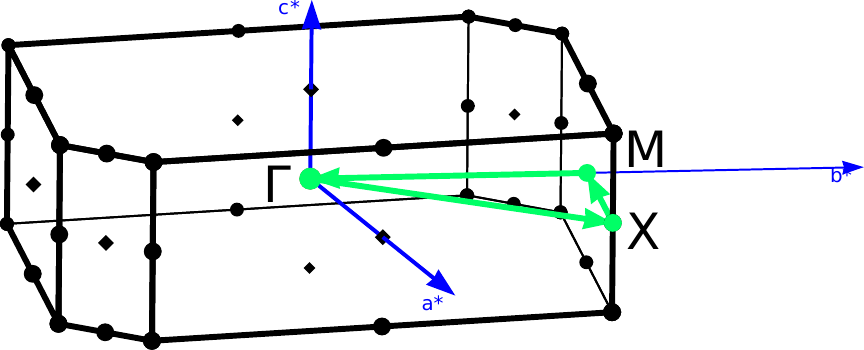}}
\end{minipage}
\begin{minipage}[h]{1\linewidth}
\begin{tabular}{p{0.32\linewidth}p{0.3\linewidth}p{0.25\linewidth}}
\centering (\textbf{a}) & \centering (\textbf{b}) & \centering (\textbf{c})
\end{tabular}
\end{minipage}
\caption{DFT (GGA) calculated band structure, the band structure with the orbital characters and the corresponding Brillouin zone (\textbf{a}--\textbf{c}) of the monoclinic $\mathrm{mlCuO}$. Zero corresponds to the Fermi level.}
\label{ris:mlCuO_monoclinic_bands_BZ}
\end{figure}

There are more surprising results for the bulk $\mathrm{CuO}$ with the monoclinic structure. Its $\mathrm{CuO}$ layers are also stretched, but to a noticeably lesser extent---by about $5.1\%$ relative to the corrugated $\mathrm{mlCuO}$. Its crystal structure has more pronounced corrugation pattern as compared to the corrugated $\mathrm{mlCuO}$ (Figure~\ref{ris:structures_bulk}c). To calculate the monoclinic $\mathrm{CuO}$ monolayer, we again made a slab+vacuum in \textit{z} direction and used no relaxation; \xadded{moreover, we did a $45^\circ$ rotation of a local coordinate system to remain consistency with the orbital convention chosen for the corrugated $\mathrm{mlCuO}$.} The final space group was 13 (\textit{P2/c}). The results are given in Figure~\ref{ris:mlCuO_monoclinic_bands_BZ}. The most surprising result is that the monoclinic $\mathrm{mlCuO}$ \xadded{has the band gap at the Fermi level, which opens because of symmetry lowering} without using DFT+U or hybrid potentials, as (but for the bulk compound) in Ref.~\cite{ekuma_electronic_2014,heinemann_band_2013}. There are mainly the \xadded{$\mathrm{Cu}$-$3d_{x^2-y^2}$} (with the hybrid $\mathrm{O}$-$2p$) states near the Fermi level similar to the corrugated $\mathrm{mlCuO}$. However, it is rather difficult to compare these results in detail with the results for the corrugated $\mathrm{mlCuO}$ (Figure~\ref{ris:bands_dos_fs}g,h) due to the different space group. Let us conclude here that a more complex corrugation pattern leads to a more complex bands structure.

\subsection{Minimal Models}

\begin{figure}[H]
\begin{minipage}[h]{0.24\linewidth}
\center{\includegraphics[width=1\linewidth]{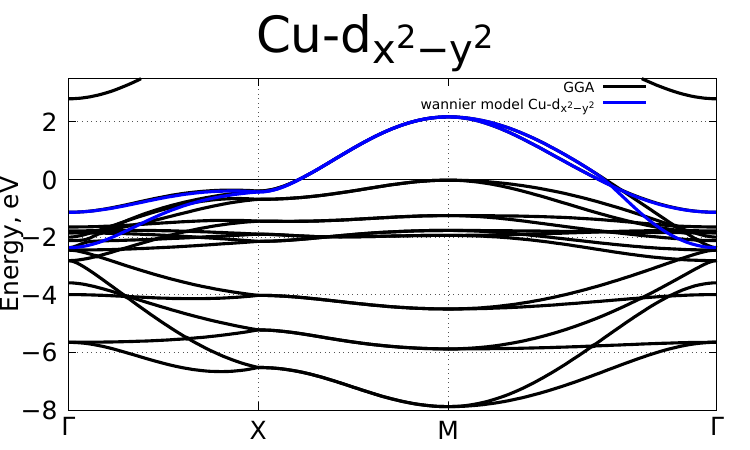}}
\end{minipage}
\begin{minipage}[h]{0.24\linewidth}
\center{\includegraphics[width=1\linewidth]{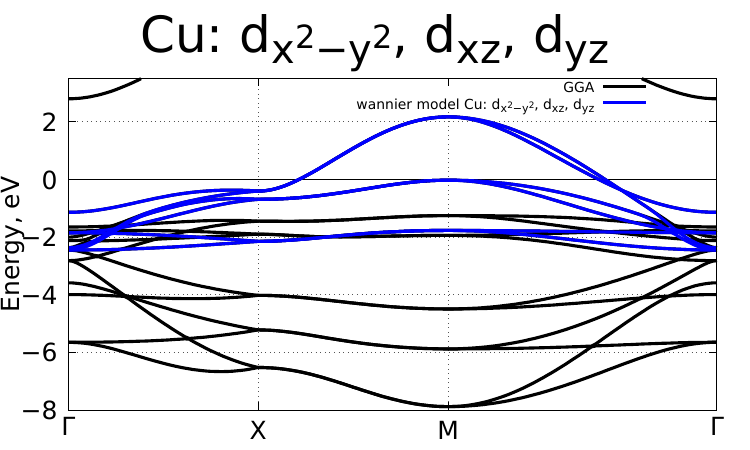}}
\end{minipage}
\begin{minipage}[h]{0.24\linewidth}
\center{\includegraphics[width=1\linewidth]{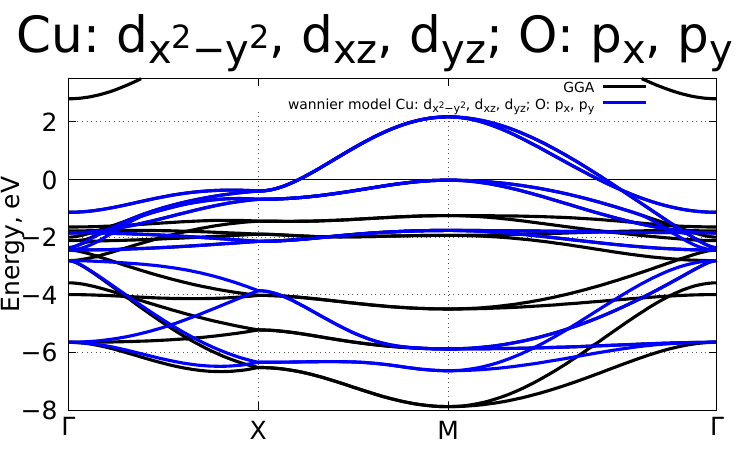}}
\end{minipage}
\begin{minipage}[h]{0.24\linewidth}
\center{\includegraphics[width=1\linewidth]{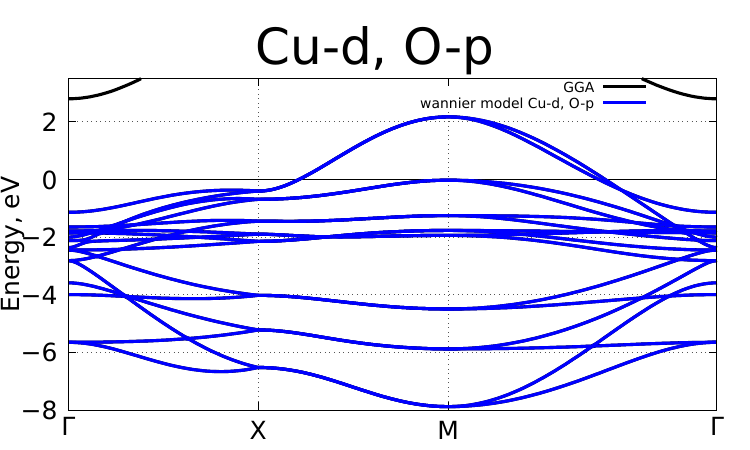}}
\end{minipage}
\begin{minipage}[h]{1\linewidth}
\begin{tabular}{p{0.22\linewidth}p{0.22\linewidth}p{0.23\linewidth}p{0.21\linewidth}}
\centering (\textbf{a}) & \centering (\textbf{b}) & \centering (\textbf{c}) & \centering (\textbf{d}) \\
\end{tabular}
\end{minipage}
\vfill
\begin{minipage}[h]{0.24\linewidth}
\center{\includegraphics[width=1\linewidth]{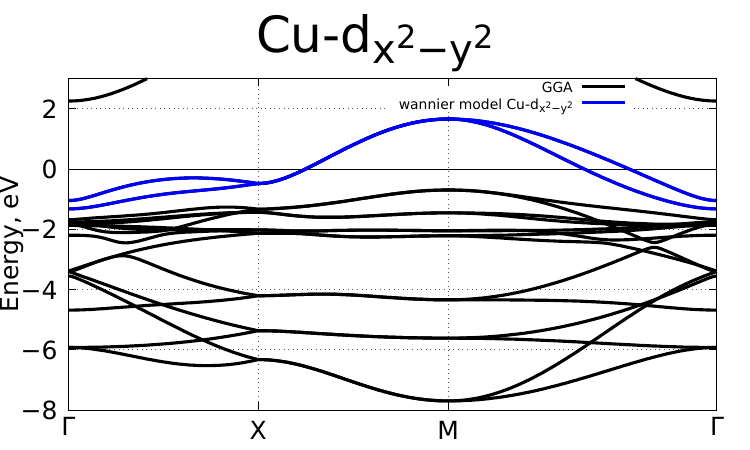}}
\end{minipage}
\hfill
\begin{minipage}[h]{0.24\linewidth}
\center{\includegraphics[width=1\linewidth]{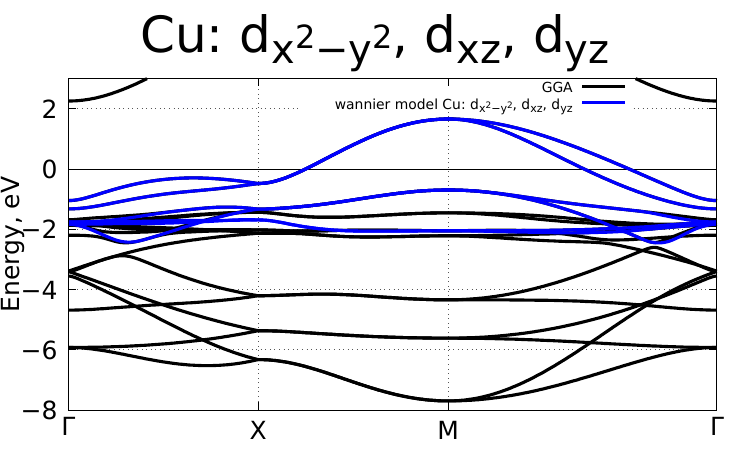}}
\end{minipage}
\hfill
\begin{minipage}[h]{0.24\linewidth}
\center{\includegraphics[width=1\linewidth]{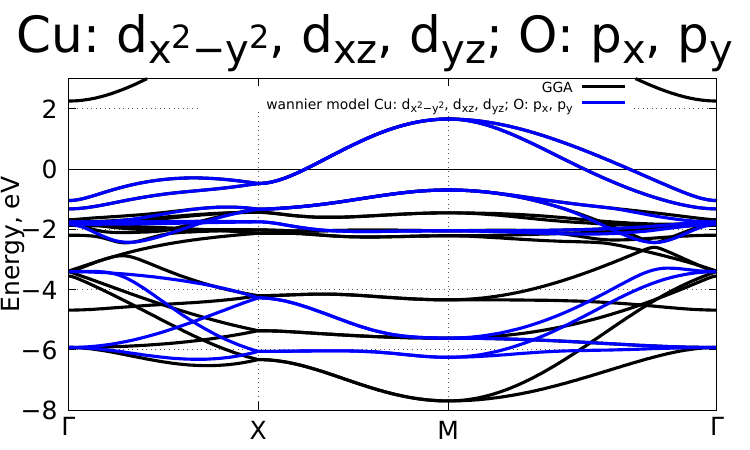}}
\end{minipage}
\hfill
\begin{minipage}[h]{0.24\linewidth}
\center{\includegraphics[width=1\linewidth]{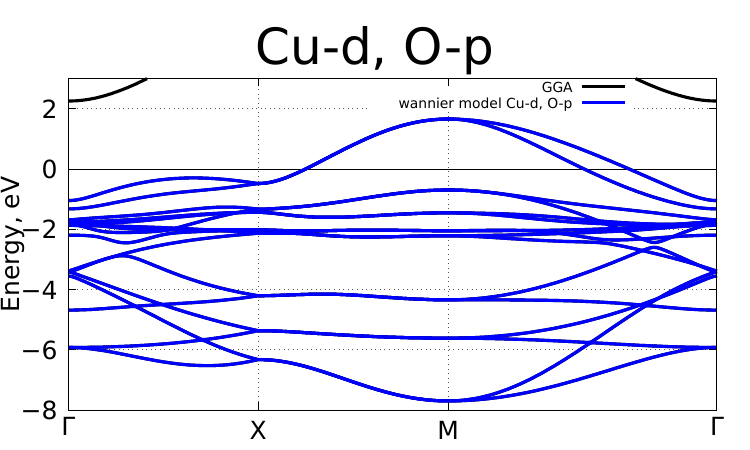}}
\end{minipage}
\begin{minipage}[h]{1\linewidth}
\begin{tabular}{p{0.22\linewidth}p{0.22\linewidth}p{0.23\linewidth}p{0.21\linewidth}}
\centering (\textbf{e}) & \centering (\textbf{f}) & \centering (\textbf{g}) & \centering (\textbf{h})
\end{tabular}
\end{minipage}
\caption{\xadded{Comparison DFT (GGA) band structure with a Wannier projected one: $\mathrm{Cu}$-$d_{x^2-y^2}$ (\textbf{a}), $\mathrm{Cu}$-$d_{x^2-y^2}, d_{xz}, d_{yz}$ (\textbf{b}), $\mathrm{Cu}$-$d_{x^2-y^2}, d_{xz}, d_{yz}$, $\mathrm{O}$-$p_x, p_y$ (\textbf{c}), $\mathrm{Cu}$-$d$, $\mathrm{O}$-$p$ (\textbf{d}) for the $\mathrm{mlCuO}$ with a doubled unit cell; same for the corrugated $\mathrm{mlCuO}$ (\textbf{e}--\textbf{h}).} Zero corresponds to the Fermi level.}
\label{ris:bands_models}
\end{figure}

The next step in our study is to reveal a good minimal model for both flat and corrugated $\mathrm{mlCuO}$. In order to do that, we constructed a set of models using the maximally localised Wannier functions (MLWF) within \texttt{wannier90} package~\cite{wannier90}. \xadded{We examined the following set of models: single-band model with Wannier projected $\mathrm{Cu}$-$d_{x^2-y^2}$, three-band model with $\mathrm{Cu}$-$d_{x^2-y^2}, d_{xz}, d_{yz}$, five-band model with $\mathrm{Cu}$-$d_{x^2-y^2}, d_{xz}, d_{yz}$, $\mathrm{O}$-$p_x, p_y$ and eight-band model with $\mathrm{Cu}$-$d$, $\mathrm{O}$-$p$.} The corresponding band structures are depicted in Figure~\ref{ris:bands_models}.
Note that there are actually twice the number of bands for the corrugated $\mathrm{mlCuO}$ and the $\mathrm{mlCuO}$ with a doubled unit cell due to unit cell doubling in contrast to the flat $\mathrm{mlCuO}$. The eight-band model resulting band structure is in the excellent agreement with GGA calculated one; in other cases, the agreement is fine at the Fermi level. Thus, the single-band \xadded{$\mathrm{Cu}$-$d_{x^2-y^2}$} model for the flat and the corrugated $\mathrm{mlCuO}$ can be used as a minimal model.

\begin{figure}[H]
\begin{minipage}[h]{0.45\linewidth}
\center{\includegraphics[scale=0.1]{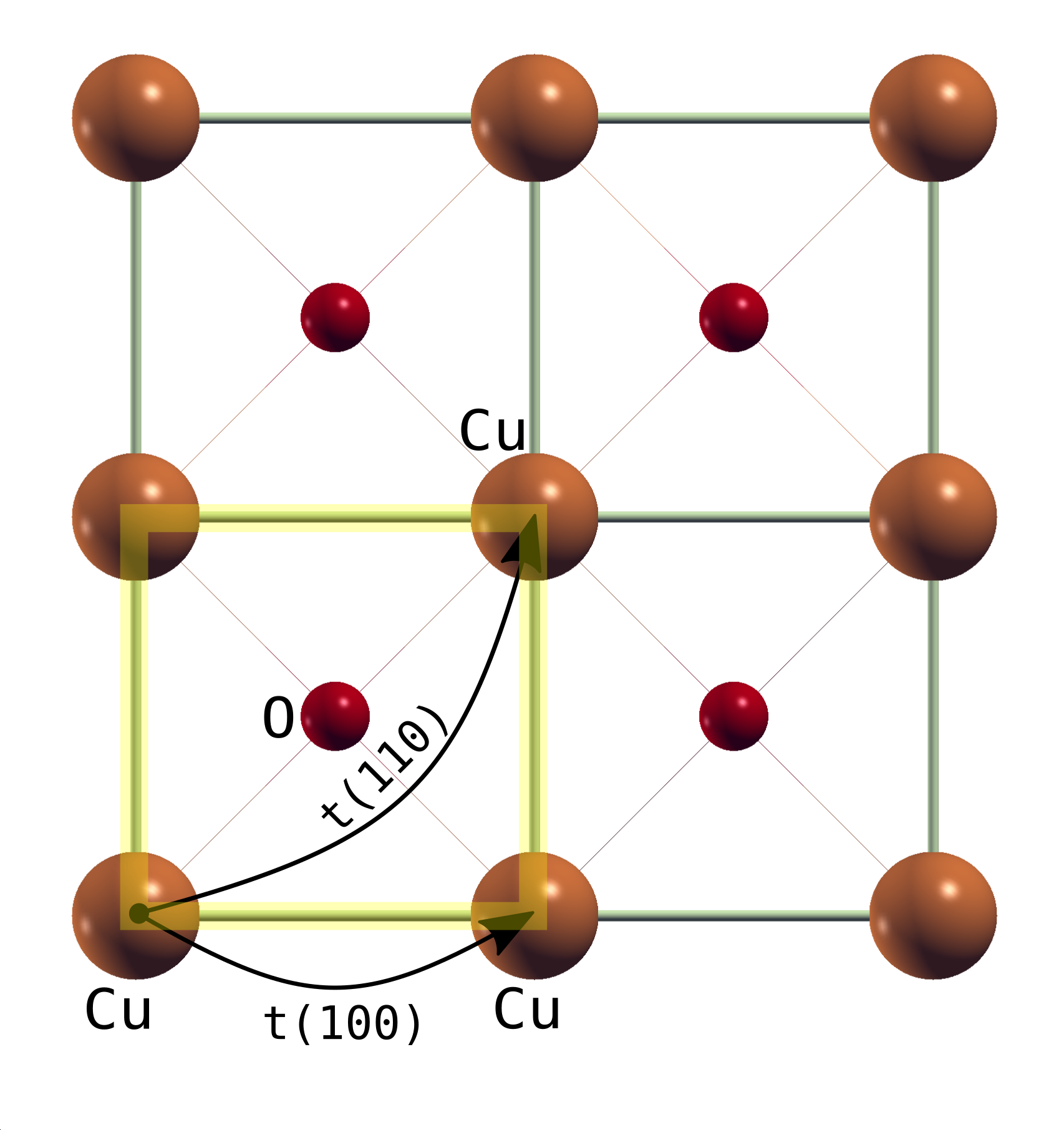}}
\end{minipage}
\hfill
\begin{minipage}[h]{0.45\linewidth}
\center{\includegraphics[scale=0.1]{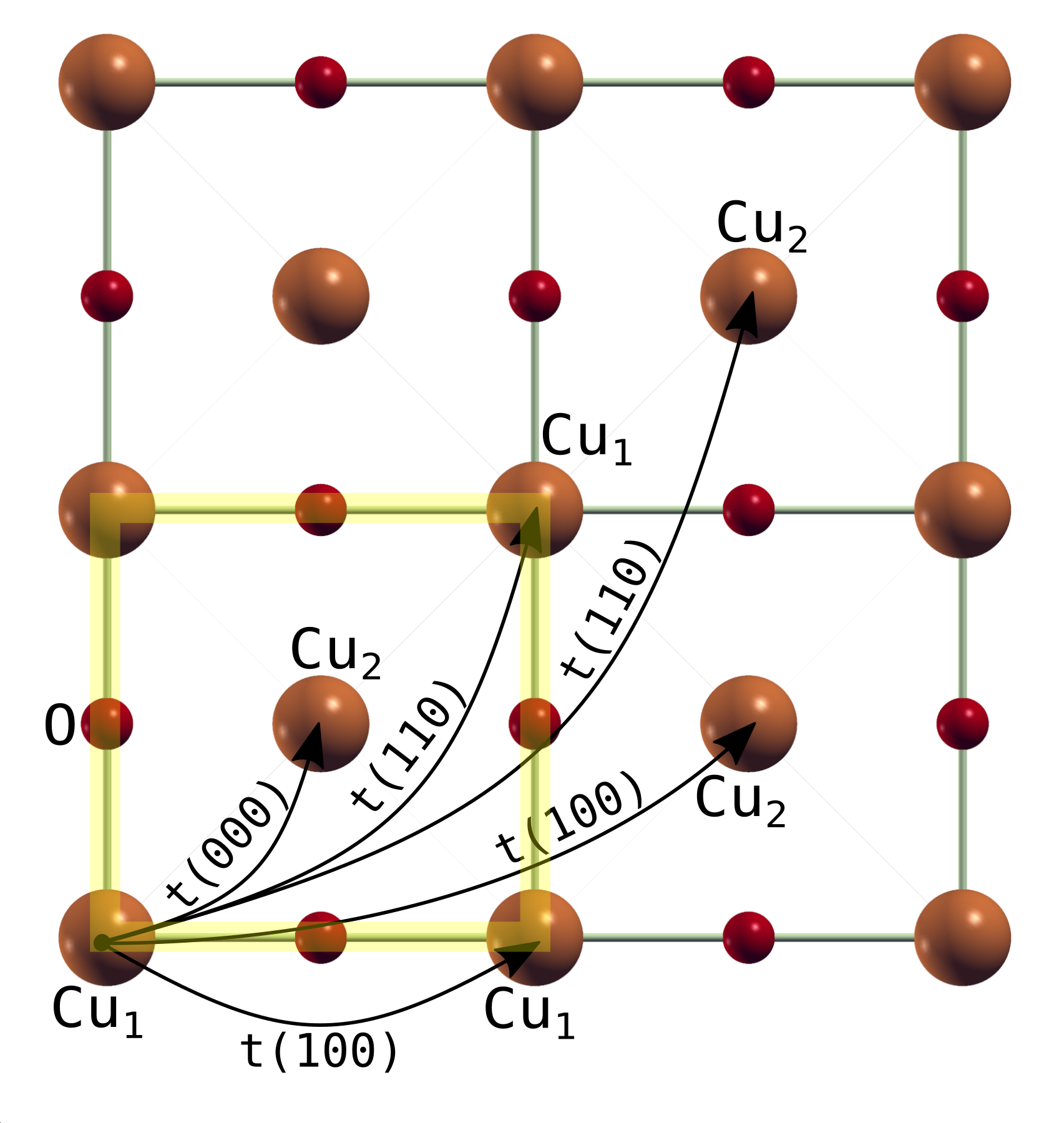}}
\end{minipage}
\caption{Hoppings schemes for the flat $\mathrm{mlCuO}$ (\textbf{left}) and the $\mathrm{mlCuO}$ with a doubled unit cell (\textbf{right}). A $2 \times 2 \times 1$ supercell is displayed for both structures. Yellow transparent square denotes a single \mbox{unit cell}.}
\label{ris:hoppings_scheme}
\end{figure}

We also present on-site energies and hopping integrals for one- and three-band models for the flat (Table~\ref{table_plain}), the corrugated $\mathrm{mlCuO}$ and the $\mathrm{mlCuO}$ with a doubled unit cell (Tables~\ref{table_dc_cor1} and~\ref{table_dc_cor2}). The corresponding hopping schemes are illustrated in Figure~\ref{ris:hoppings_scheme}. We used hoppings up to the second coordination sphere for the flat $\mathrm{mlCuO}$ and up to the fifth coordination sphere for the corrugated one. These numbers of neighbors are the minimum required to obtain good agreement between the model Hamiltonian band structure and the initial one. We also attach the values of the Hamiltonian matrix elements in a real space for all the models (Figure~\ref{ris:bands_models}) as machine-readable data files to supplementary materials. So, depending on a task, an interested reader can use the Hamiltonian of the \mbox{appropriate complexity.}

\vspace{-6pt}


\section{Conclusions}

We investigated the electronic properties of the flat and the corrugated $\mathrm{CuO}$ \linebreak \mbox{monolayers---DOS}, the band structures and the Fermi surfaces. The flat $\mathrm{mlCuO}$ is similar to the bulk $\mathrm{CuO}$ with the monoclinic crystal structure and typical $\mathrm{Cu}$-based HTSC, e.g., $\mathrm{La_2CuO_4}$ and has the $\mathrm{Cu}$-$3d_{x^2-y^2}$ states at the Fermi level (with a small addition of the hybrid $\mathrm{O}$-$2p$ states). There is a second band just below the Fermi level---that it is only 0.02~eV lower. It includes the $\mathrm{Cu}$-$3d_{xz,yz}$ states hybridized with $\mathrm{O}$-$2p_z$.

The corrugation effect leads to \xreplaced{}{the replacement of the $\mathrm{Cu}$-$3d_{x^2-y^2}$ states with the $\mathrm{Cu}$-$3d_{xy}$ ones,} a significant shift of the $\mathrm{Cu}$-$3d_{xz,yz}$ bands to $-0.7$~eV and a degeneracy lifting for the \xadded{$\mathrm{Cu}$-$3d_{x^2-y^2}$} bands. The corrugated $\mathrm{mlCuO}$ is energetically more favorable than the flat one by $0.07$~eV and is more likely formed as a topmost layer on some surfaces.

It is possible to create a topological Lifshitz transition via lattice strain: for the flat $\mathrm{mlCuO}$, a slight stretching of the lattice parameter $\Delta a = 0.7\%$ already leads to the appearance of a hole pocket around the $X$ point; this effect for the corrugated $\mathrm{mlCuO}$ occurs at $\Delta a = 35\%$. It is interesting to note the presence of what seems to be a rather flat band in the $\Gamma - X$ direction near the Fermi level. Probably, it will be possible to create it using a reasonable strain and a hole doping.

There is a significant mismatch in the lattice parameters of the considered $\mathrm{CuO}$ monolayers and the known bulk $\mathrm{CuO}$ systems (with the cubic and the monoclinic crystal structures). We conclude that $\mathrm{CuO}$ layers in the bulk $\mathrm{CuO}$ are stretched relative to the monolayer systems---by about $11.5$\% (cubic bulk vs. flat monolayer) and by $5.1$\% (monoclinic bulk vs. corrugated monolayer). The monolayer made from experimental bulk $\mathrm{CuO}$ with the monoclinic crystal structure turns out to \xadded{have the band gap} in our DFT calculation without using DFT+U or hybrid potentials. Clearly, the reason for this is the more complex corrugation patterns of its crystal structure.

We also suggested a set of minimal models for the flat and the corrugated $\mathrm{CuO}$ monolayers. The simplest model includes only the \xadded{$\mathrm{Cu}$-$d_{x^2-y^2}$} states and agrees well with the GGA calculated band structure at the Fermi level (the latter is also correct for the other models). For the one- and three-band models, we obtained the values of the corresponding Hamiltonian matrix elements in a real space; therefore, depending on the needs, the Hamiltonian of the appropriate complexity can be used.

Before proceeding to more complex systems, such as a $\mathrm{CuO}$ monolayer on a substrate or a $\mathrm{CuO}$ monolayer as an interface, it was necessary to perform calculations for the original system and understand the  features of its electronic structure. So, the next step might be the investigation of such complex systems or using more advanced methods such as DFT+DMFT.

\vspace{6pt}


 \supplementary{The following supporting information can be downloaded at: \linksupplementary{s1}.}

\authorcontributions{Conceptualization, I.A.N., M.M.K. and S.G.O.; calculations, A.A.S. and L.V.B.; writing, A.A.S. and I.A.N.; review and editing, I.A.M., M.M.K. and S.G.O.; funding acquisition, I.A.N. All authors have read and agreed to the published version of the manuscript.}

\funding{{This} 
 research was in part funded by the Russian Science Foundation grant number 23-22-00372.}

\acknowledgments{\xadded{L.V.B. would like to thank the Irkutsk Supercomputer Center of SB RAS for providing the access to HPC-cluster «Akademik V.M. Matrosov» (Irkutsk Supercomputer Center of SB RAS, Irkutsk: ISDCT SB RAS; \url{http://hpc.icc.ru}, accessed 21 November 2022).}}

\conflictsofinterest{The authors declare no conflict of interest. The funders had no role in the design of the study; in the collection, analyses, or interpretation of data; in the writing of the manuscript; or in the decision to publish the~results.}



%

\appendixtitles{no} 
\appendixstart
\appendix
\section[\appendixname~\thesection]{}

We attach the values of the Hamiltonian matrix elements of all the models considered in this work as supplementary materials. These are machine-readable data files with self-explanatory names and some comments.

\begin{table}[H]
\caption{An on-site energies and hoppings in meV for the flat $\mathrm{mlCuO}$ (one- and {three-band models}).}

\label{table_plain}

 \newcolumntype{M}[1]{>{\centering\arraybackslash}m{#1}}
\begin{tabularx}{\textwidth}{M{3CM}CCCCCC}
\toprule
 &   \multicolumn{3}{c}{\boldmath{$\mathrm{Cu}$-$d_{x^2-y^2}$}} & \multicolumn{3}{c}{\boldmath{$\mathrm{Cu}$-$d_{x^2-y^2,xz,yz}$}} \\
\cmidrule{2-7}
\boldmath{$m^{\prime},m$} &  \boldmath{$\varepsilon(000)$} & \boldmath{$t(100)$} & \boldmath{$t(110)$} &  \boldmath{$\varepsilon(000)$} & \boldmath{$t(100)$} & \boldmath{$t(110)$} \\
\midrule
$x^2-y^2,x^2-y^2$ & 	$126$	&	$-69$	&	$-485$	& 	$169$	&	$-43$	&	$-461$\\
$xz,xz$ & 			&			&			& 	$-1417$	&	$-227$	&	$-139$\\
$yz,yz$ & 			&			&			& 	$-1417$	&	$226$	&	$-139$\\
$xz,yz$ & 			&			&			& 	$0$		&	$0$		&	$-191$\\
\bottomrule
\end{tabularx}
\end{table}

\vspace{-6pt}

\begin{table}[H]
\caption{An on-site energies and hoppings in meV for the doubled unit cell and corrugated $\mathrm{mlCuO}$ (one-band model).}
\label{table_dc_cor1}

 \newcolumntype{M}[1]{>{\centering\arraybackslash}m{#1}}
\begin{tabularx}{\textwidth}{M{3.5CM}CCCCCC}
\toprule
   & \multicolumn{6}{c}{\boldmath{$\mathrm{Cu_{1,2}}$-$d_{x^2-y^2}$}} \\\cmidrule{2-7}
  & \multicolumn{3}{c}{\textbf{Doubled u.c.}} &   \multicolumn{3}{c}{\textbf{Corrugated}}
\\ \cmidrule{2-7}

 \boldmath{$m^{\prime},m$}& \boldmath{$\varepsilon(000)$} & \boldmath{$t(100)$} & \boldmath{$t(110)$} & \boldmath{$\varepsilon(000)$} & \boldmath{$t(100)$} & \boldmath{$t(110)$} \\ \midrule

$x^2-y^2_\mathrm{Cu_1},x^2-y^2_\mathrm{Cu_1}$ & 	$124$		&	$-486$	&	$73$	& 	$161$	&	$-351$	&	$86$\\
$x^2-y^2_\mathrm{Cu_1},x^2-y^2_\mathrm{Cu_2}$ & 	$-70$		&	$-43$	&	$13$	& 	$5$		&	$-19$	&	$-19$\\
\bottomrule
\end{tabularx}%


\end{table}

\begin{table}[H]
\caption{An on-site energies and hoppings in meV for the doubled unit cell and corrugated $\mathrm{mlCuO}$ (three-band model).}
\label{table_dc_cor2}

 \newcolumntype{M}[1]{>{\centering\arraybackslash}m{#1}}
\begin{tabularx}{\textwidth}{M{3.5CM}CCCCCC}
\toprule
 &   \multicolumn{6}{c}{\boldmath{$\mathrm{Cu_{1,2}}$-$d_{x^2-y^2,xz,yz}$}} \\\cmidrule{2-7}

&   \multicolumn{3}{c}{\textbf{Doubled u.c.}} &   \multicolumn{3}{c}{\textbf{Corrugated}} \\ \cmidrule{2-7}

\boldmath{$m^{\prime},m$} & \boldmath{$\varepsilon(000)$} & \boldmath{$t(100)$} & \boldmath{$t(110)$} &  \boldmath{$\varepsilon(000)$} & \boldmath{$t(100)$} & \boldmath{$t(110)$} \\
\midrule
$x^2-y^2_\mathrm{Cu_1},x^2-y^2_\mathrm{Cu_1}$ & 	$167$	& $-462$	&	$82$	& 	$24$	&	$-391$	&	$85$\\
$x^2-y^2_\mathrm{Cu_1},x^2-y^2_\mathrm{Cu_2}$ & 	$-44$	& $-36$		&	$10$	& 	$-39$	&	$-19$	&	$-19$\\
\midrule
$x^2-y^2_\mathrm{Cu_1},xz_\mathrm{Cu_1}$ &  $0$		&	$0$		&	$0$		&  $0$		&	$-278$	&	$30$\\
$x^2-y^2_\mathrm{Cu_1},xz_\mathrm{Cu_2}$ &  $0$		&	$0$		&	$0$		&  $41$		&	$31$	&	$31$\\
$x^2-y^2_\mathrm{Cu_2},xz_\mathrm{Cu_1}$ &  $0$		&	$0$		&	$0$		&  $41$		&	$-41$	&	$7$\\
$x^2-y^2_\mathrm{Cu_1},yz_\mathrm{Cu_2}$ &  $0$		&	$0$		&	$0$		&  $-20$		&	$-34$	&	$34$\\
\midrule
$xz_\mathrm{Cu_1},xz_\mathrm{Cu_1}$ &  $-1417$	&	$-329$	&	$-8$	& 	$-1430$	&	$-128$	&	$-40$\\
$xz_\mathrm{Cu_1},xz_\mathrm{Cu_2}$ & $18$		&	$41$	&	$5$		& 	$0$		&	$14$	&	$14$\\
\midrule
$yz_\mathrm{Cu_1},yz_\mathrm{Cu_1}$ &  $-1417$	&	$53$	&	$-8$	& 	$-1439$	&	$43$	&	$-3$\\
$yz_\mathrm{Cu_1},yz_\mathrm{Cu_2}$ &  $18$		&	$-12$	&	$5$		& 	$-6$	&	$4$		&	$4$\\
\midrule
$xz_\mathrm{Cu_1},yz_\mathrm{Cu_1}$ &  $0$		& $0$		&	$47$	& 	$0$		&	$0$		&	$28$\\
$xz_\mathrm{Cu_1},yz_\mathrm{Cu_2}$ &  $-246$		& $-21$		&	$-214$	& 	$-206$	&	$2$		&	$-2$\\
$xz_\mathrm{Cu_2},yz_\mathrm{Cu_1}$ &  $-246$		& $246$		&	$-246$	& 	$206$	&	$-206$	&	$-24$\\
\bottomrule
\end{tabularx}%

\end{table}

\begin{adjustwidth}{-\extralength}{0cm}
\printendnotes[custom] 

\reftitle{References}

\PublishersNote{}
\end{adjustwidth}

\begin{thebibliography}{999}

\bibitem[Anisimov \em{et~al.}(1991)Anisimov, Zaanen, and
  Andersen]{anisimov_band_1991}
Anisimov, V.I.; Zaanen, J.; Andersen, O.K.
\newblock Band theory and Mott insulators: Hubbard U instead of Stoner I.
\newblock {\em Phys. Rev. B} {\bf 1991}, {\em 44},~943--954.
  {{https://doi.org/10.1103/PhysRevB.44.943}}.

\bibitem[Ruiz \em{et~al.}(1997)Ruiz, Alvarez, Alemany, and
  Evarestov]{ruiz_electronic_1997}
Ruiz, E.; Alvarez, S.; Alemany, P.; Evarestov, R.A.
\newblock Electronic structure and properties of $\mathrm{Cu_2O}$.
\newblock {\em Phys. Rev. B} {\bf 1997}, {\em 56},~7189--7196.
  {{https://doi.org/10.1103/PhysRevB.56.7189}}.

\bibitem[Ghijsen \em{et~al.}(1998)Ghijsen, Tjeng, van Elp, Eskes, Westerink,
  Sawatzky, and Czyzyk]{ghijsen_electronic_1988}
Ghijsen, J.; Tjeng, L.H.; van Elp, J.; Eskes, H.; Westerink, J.; Sawatzky,
  G.A.; Czyzyk, M.T.
\newblock Electronic structure of $\mathrm{Cu_2O}$ and $\mathrm{CuO}$.
\newblock {\em Phys. Rev. B} {\bf 1998}, {\em 38},~11322--11330.
  {{https://doi.org/10.1103/PhysRevB.38.11322}}.

\bibitem[Heinemann \em{et~al.}(2013)Heinemann, Eifert, and
  Heiliger]{heinemann_band_2013}
Heinemann, M.; Eifert, B.; Heiliger, C.
\newblock Band structure and phase stability of the copper oxides
  $\mathrm{Cu_2O}$, $\mathrm{CuO}$, and $\mathrm{Cu_4O_3}$.
\newblock {\em Phys. Rev. B} {\bf 2013}, {\em 87},~115111.
  {{https://doi.org/10.1103/PhysRevB.87.115111}}.

\bibitem[Pickett(1989)]{pickett_electronic_1989}
Pickett, W.E.
\newblock Electronic structure of the high-temperature oxide superconductors.
\newblock {\em Rev. Mod. Phys.} {\bf 1989}, {\em 61},~433--512.
  {{https://doi.org/10.1103/RevModPhys.61.433}}.

\bibitem[Reitz and Solomon(1998)]{reitz_propylene_1998}
Reitz, J.B.; Solomon, E.I.
\newblock Propylene Oxidation on Copper Oxide Surfaces: Electronic and
  Geometric Contributions to Reactivity and Selectivity.
\newblock {\em J. Am. Chem. Soc.} {\bf 1998}, {\em
  120},~11467--11478.
  {{https://doi.org/10.1021/ja981579s}}.

\bibitem[Nakaoka \em{et~al.}(2004)Nakaoka, Ueyama, and
  Ogura]{nakaoka_photoelectrochemical_2004}
Nakaoka, K.; Ueyama, J.; Ogura, K.
\newblock Photoelectrochemical Behavior of Electrodeposited $\mathrm{CuO}$ and
  $\mathrm{Cu_2O}$ Thin Films on Conducting Substrates.
\newblock {\em J. Electrochem. Soc.} {\bf 2004}, {\em
  151},~C661.
  {{https://doi.org/10.1149/1.1789155}}.

\bibitem[Kim \em{et~al.}(2013)Kim, Ahn, Lee, Kwon, Hwang, Lee, and
  Cho]{kim_p-channel_2013}
Kim, S.Y.; Ahn, C.H.; Lee, J.H.; Kwon, Y.H.; Hwang, S.; Lee, J.Y.; Cho, H.K.
\newblock p-Channel Oxide Thin Film Transistors Using Solution-Processed Copper
  Oxide.
\newblock {\em {ACS} Appl. Mater. Interfaces} {\bf 2013}, {\em
  5},~2417--2421.
  {{https://doi.org/10.1021/am302251s}}.

\bibitem[Åsbrink and Norrby(1970)]{asbrink_refinement_1970}
Åsbrink, S.; Norrby, L.J.
\newblock A refinement of the crystal structure of copper({II}) oxide with a
  discussion of some exceptional e.s.d.'s.
\newblock {\em Acta Crystallogr. Sect. B} {\bf 1970}, {\em 26},~8--15.
  {{https://doi.org/10.1107/S0567740870001838}}.

\bibitem[Schmahl and Eikerling(1968)]{schmahl_uber_1968}
Schmahl, N.G.; Eikerling, G.F.
\newblock Über Kryptomodifikationen des Cu({II})-Oxids.
\newblock {\em Z. Für Phys. Chem.} {\bf 1968}, {\em
  62},~268--279.
 {{https://doi.org/10.1524/zpch.1968.62.5\_6.268}}.

\bibitem[Ekuma \em{et~al.}(2014)Ekuma, Anisimov, Moreno, and
  Jarrell]{ekuma_electronic_2014}
Ekuma, C.; Anisimov, V.; Moreno, J.; Jarrell, M.
\newblock Electronic structure and spectra of $\mathrm{CuO}$.
\newblock {\em  Eur. Phys. J. B} {\bf 2014}, {\em 87},~23.
\newblock {{https://doi.org/10.1140/epjb/e2013-40949-5}}.

\bibitem[Wu \em{et~al.}(2006)Wu, Zhang, and Tao]{wu_mathrmlsdamathrmu_2006}
Wu, D.; Zhang, Q.; Tao, M.
\newblock LSDA+U study of cupric oxide: Electronic structure and native point
  defects.
\newblock {\em Phys. Rev. B} {\bf 2006}, {\em 73},~235206.
  {{https://doi.org/10.1103/PhysRevB.73.235206}}.

\bibitem[Nolan and Elliott(2006)]{nolan_p-type_2006}
Nolan, M.; Elliott, S.D.
\newblock The p-type conduction mechanism in $\mathrm{Cu_2O}$: A first
  principles study.
\newblock {\em Phys. Chem. Chem. Phys.} {\bf 2006}, {\em
  8},~5350--5358.
  {{https://doi.org/10.1039/B611969G}}.

\bibitem[Cao \em{et~al.}(2018)Cao, Zhou, Yu, and Zhou]{cao_dft_2018}
Cao, H.; Zhou, Z.; Yu, J.; Zhou, X.
\newblock DFT study on structural, electronic, and optical properties of cubic
  and monoclinic $\mathrm{CuO}$.
\newblock {\em J. Comput. Electron.} {\bf 2018}, {\em
  17},~21--28.
\newblock {{https://doi.org/10.1007/s10825-017-1057-9}}.

\bibitem[Grant(2008)]{grant_electronic_2008}
Grant, P.M.
\newblock Electronic properties of rocksalt copper monoxide: a proxy structure
  for high temperature superconductivity.
\newblock {\em J. Phys. Conf. Ser.} {\bf 2008}, {\em
  129},~012042.
  {{https://doi.org/10.1088/1742-6596/129/1/012042}}.

\bibitem[Cipriano \em{et~al.}(2020)Cipriano, Di~Liberto, Tosoni, and
  Pacchioni]{cipriano_band_2020}
Cipriano, L.A.; Di~Liberto, G.; Tosoni, S.; Pacchioni, G.
\newblock Band Gap in Magnetic Insulators from a Charge Transition Level
  Approach.
\newblock {\em J. Chem. Theory Comput.} {\bf 2020}, {\em
  16},~3786--3798.
  {{https://doi.org/10.1021/acs.jctc.0c00134}}.

\bibitem[Yazdani and Barakati(2021)]{yazdani_first-principles_2021}
Yazdani, A.; Barakati, B.
\newblock A first-principles study on electronic structure and crystal field
  effect of layered La\textsubscript{2}CuO\textsubscript{4} as composed of {CuO}\textsubscript{2} and La\textsubscript{2}O\textsubscript{2} monolayers.
\newblock {\em Phys. E Low-Dimens. Syst. Nanostructures} {\bf
  2021}, {\em 125},~114395.
\newblock {{https://doi.org/10.1016/
j.physe.2020.114395}}.

\bibitem[Yin \em{et~al.}(2016)Yin, Zhang, Zhou, Sun, Chisholm, Pantelides, and
  Zhou]{yin_unsupported_2016}
Yin, K.; Zhang, Y.Y.; Zhou, Y.; Sun, L.; Chisholm, M.F.; Pantelides, S.T.;
  Zhou, W.
\newblock Unsupported single-atom-thick copper oxide monolayers.
\newblock {\em 2D Mater.} {\bf 2016}, {\em 4},~011001.
  {{https://doi.org/10.1088/2053-1583/4/1/011001}}.

\bibitem[Kano \em{et~al.}(2017)Kano, G.~Kvashnin, Sakai, A.~Chernozatonskii,
  B.~Sorokin, Hashimoto, and Takeguchi]{kano_one-atom-thick_2017}
Kano, E.; G.~Kvashnin, D.; Sakai, S.; A.~Chernozatonskii, L.; B.~Sorokin, P.;
  Hashimoto, A.; Takeguchi, M.
\newblock One-atom-thick 2D copper oxide clusters on graphene.
\newblock {\em Nanoscale} {\bf 2017}, {\em 9},~3980--3985.
  {{https://doi.org/10.1039/C6NR06874J}}.

\bibitem[Kvashnin \em{et~al.}(2019)Kvashnin, Kvashnin, Kano, Hashimoto,
  Takeguchi, Naramoto, Sakai, and Sorokin]{kvashnin_two-dimensional_2019}
Kvashnin, D.G.; Kvashnin, A.G.; Kano, E.; Hashimoto, A.; Takeguchi, M.;
  Naramoto, H.; Sakai, S.; Sorokin, P.B.
\newblock Two-Dimensional $\mathrm{CuO}$ Inside the Supportive Bilayer Graphene
  Matrix.
\newblock {\em  J. Phys. Chem. C} {\bf 2019}, {\em
  123},~17459--17465.
  {{https://doi.org/10.1021/acs.jpcc.9b05353}}.

\bibitem[Blaha \em{et~al.}(2020)Blaha, Schwarz, Tran, Laskowski, Madsen, and
  Marks]{wien2k}
Blaha, P.; Schwarz, K.; Tran, F.; Laskowski, R.; Madsen, G.K.H.; Marks, L.D.
\newblock WIEN2k: An APW+lo program for calculating the properties of solids.
\newblock {\em  J. Chem. Phys.} {\bf 2020}, {\em 152},~074101.
  {{https://doi.org/10.1063/1.5143061}}.

\bibitem[Perdew \em{et~al.}(1996)Perdew, Burke, and Ernzerhof]{pbe}
Perdew, J.P.; Burke, K.; Ernzerhof, M.
\newblock Generalized Gradient Approximation Made Simple.
\newblock {\em Phys. Rev. Lett.} {\bf 1996}, {\em 77},~3865--3868.
\newblock {{https://doi.org/10.1103/PhysRevLett.77.3865}}.

\bibitem[Monkhorst and Pack(1976)]{monkhorst1976special}
Monkhorst, H.J.; Pack, J.D.
\newblock Special points for Brillouin-zone integrations.
\newblock {\em Phys. Rev. B} {\bf 1976}, {\em 13},~5188.

\bibitem[Pizzi \em{et~al.}(2020)Pizzi, Vitale, Arita, Blügel, Freimuth,
  Géranton, Gibertini, Gresch, Johnson, Koretsune, Ibañez-Azpiroz, Lee, Lihm,
  Marchand, Marrazzo, Mokrousov, Mustafa, Nohara, Nomura, Paulatto, Poncé,
  Ponweiser, Qiao, Thöle, Tsirkin, Wierzbowska, Marzari, Vanderbilt, Souza,
  Mostofi, and Yates]{wannier90}
Pizzi, G.; Vitale, V.; Arita, R.; Blügel, S.; Freimuth, F.; Géranton, G.;
  Gibertini, M.; Gresch, D.; Johnson, C.; Koretsune, T.;  et~al.
\newblock Wannier90 as a community code: new features and applications.
\newblock {\em J. Phys. Condens. Matter.} {\bf 2020}, {\em
  32},~165902.
  {{https://doi.org/10.1088/1361-648X/ab51ff}}.

\end{thebibliography}
\end{document}